\documentclass[pageno]{jpaper}

\usepackage[normalem]{ulem}

\usepackage{mathptmx} % This is Times font
\usepackage{amsmath}
\usepackage{amsfonts}
\usepackage{amssymb}
\DeclareMathOperator*{\argmax}{arg\,max}

\usepackage{graphicx}
\usepackage{multirow}
\usepackage{fancyhdr}
\usepackage[sort,nocompress]{cite}
\usepackage[keeplastbox]{flushend}
\usepackage{xcolor}
\usepackage{physics}
\usepackage[ruled,vlined,linesnumbered,resetcount,noend]{algorithm2e}
\usepackage{setspace}
\usepackage{todonotes}

\usepackage{syntax}
\usepackage{array}
\newcolumntype{?}{!{\vrule width 1pt}}

\usepackage[numbers,sort&compress]{natbib}

\newcommand{\myCompilerName}{Paulihedral}
\newcommand{\myCompilerNameSpace}{Paulihedral }
\newcommand{\QSK}{quantum simulation kernel}
\newcommand{\QSKSpace}{quantum simulation kernel }
\newcommand{\ps}{Pauli string}
\newcommand{\psSpace}{Pauli string }
\definecolor{myorange}{HTML}{B25A00}

\usepackage{amsmath}
\usepackage{amssymb}
\usepackage{amsthm}
\usepackage{stmaryrd}

\def\>{\ensuremath{\rangle}}
\def\<{\ensuremath{\langle}}

\newcommand {\sem}[1] {\llbracket#1\rrbracket}

\begin{document}

\title{\myCompilerName: A Generalized Block-Wise Compiler Optimization Framework For Quantum Simulation Kernels}
\author{$^{1}$Gushu Li, $^{1}$Anbang Wu, $^{2}$Yunong Shi, $^{3}$Ali Javadi-Abhari, $^{1}$Yufei Ding, $^{1}$Yuan Xie \\
$^{1}$University of California, Santa Barbara, CA, USA\\
$^{2}$Amazon Braket, NY, USA\\
$^{3}$IBM Quantum, NY, USA}

\date{}
\maketitle

\thispagestyle{empty}

\begin{abstract}
The quantum simulation kernel is an important subroutine appearing as a very long gate sequence in many quantum programs.
In this paper, we propose \myCompilerName, a block-wise compiler framework that can deeply optimize this subroutine by exploiting high-level program structure and optimization opportunities.
\myCompilerNameSpace first employs a new Pauli intermediate representation that can maintain the high-level semantics and constraints in quantum simulation kernels. 
This naturally enables new large-scale optimizations that are hard to implement at the low gate-level. 
%, are enabled at.
%to express and manage quantum simulation kernels with high-level semantics maintained.
%Several new large-scale optimizations, which are hard to implement at the low gate-level, are enabled at.
%, enabling new optimizations that are hard to implement at the gate-level.
%We then propose several novel optimization passes, all of which are organized at the block-level.
In particular, we propose two technology-independent instruction scheduling passes, and two technology-dependent code optimization passes which reconcile the circuit synthesis, gate cancellation, and qubit mapping stages of the compiler.
Experimental results show that \myCompilerNameSpace can outperform state-of-the-art compiler infrastructures in a wide-range of applications on both near-term superconducting quantum processors and future fault-tolerant quantum computers.

%The quantum simulation principle is widely used in the quantum algorithm design and thus a subroutine called the quantum simulation kernel appears frequently in many quantum programs. 
%This subroutine usually results in a very long post-compilation gate sequence, since today's compiler optimizations for this subroutine are far from optimal and many high-level algorithmic optimization opportunities are not exploited.
%hard to be leveraged in gate-based compiler infrastructure.
%\YS{(because not optimal, thus long sequence), could mention they are not optimal because they didn't respect the global information}.
%In this paper, we propose \myCompilerName, a block-wise compiler framework that can deeply optimize the quantum simulation kernels.
%\myCompilerNameSpace employs a new Pauli intermediate representation to express and manage quantum simulation kernels with high-level semantics maintained, enabling new optimizations that are hard to be implemented at the gate-level.
%We then propose several novel optimization passes, all of which are organized at the block-level.
%In particular, we have two technology-independent instruction scheduling passes, and two technology-dependent code optimization passes which reconcile the circuit synthesis, gate cancellation, and qubit mapping stages of the compiler.
%Experimental results show that \myCompilerNameSpace can outperform state-of-the-art compiler infrastructures in a wide-range of applications on both near-term superconducting quantum processors and future fault-tolerant quantum computers.
%\todo{}

\end{abstract}

\section{Introduction}\label{sec:introduction}

One of the most important quantum algorithm design principles is quantum simulation (or Hamiltonian simulation).
Simulating a quantum physical system, which motivated Feynman's proposal
%was the original motivation \YS{(motivated Richard Feynman..)} when Richard Feynman proposed 
to build a quantum computer~\cite{feynman1982simulating}, %back in 1982 \YS{no need to mention year}, 
is by itself an important application of quantum computing% and then the corresponding quantum algorithms were invented
~\cite{lloyd1996universal, abrams1999quantum}.
Later, the idea of quantum simulation was extended to quantum algorithms for other applications, e.g., linear systems~\cite{harrow2009quantum}, quantum principal component analysis~\cite{lloyd2014quantum}, and quantum support vector machine~\cite{rebentrost2014quantum}. 
These algorithms involve simulating an artificial quantum system crafted based on the target problem.
In recently developed variational quantum algorithms for near-term quantum computers (e.g., VQE for chemistry~\cite{peruzzo2014variational} and QAOA for combinatorial optimization~\cite{farhi2014quantum}), the program structures are also inspired by
%the circuit structures of the ansatzes (i.e., parameterized quantum programs) are also inspired by 
%this principle.
the simulation principle.

Because the quantum simulation principle is shared among many algorithms, one subroutine, which we term the \textit{\QSK} in this paper, appears frequently in quantum programs.
This kernel is to implement the operator (controlled-)${\rm exp}({\rm i}Ht)$ where $H$ is the Hamiltonian of the simulated system
%the simulated physical system \YS{ target problem? system of interest? Not necessarily physical}
and $t \in \mathbb{R}$ is system evolution time. % representing time \YS{(just time is ok}). 
Since it is hard in general to directly compile ${\rm exp}({\rm i}Ht)$ into executable single- and two-qubit gates,
%In practice 
%To implement this operator, 
a compiler usually decomposes $H$ into the sum of local Hamiltonians~\cite{lloyd1996universal} (simulation of which can be easily compiled to basic gates) and then synthesize them one-by-one.
%and approximates a small time step ${\rm exp}({\rm i}H\Delta t)$ using the Trotter formula~\cite{trotter1959product}. 
%\YS{1. the logis is more like: most practical Hamiltonians can be written as a sum of local Hamiltonians 2. should mention Trotter formula is an approximation}
%Then the operator ${\rm exp}({\rm i}H\Delta t)$ will be repeated many times to finally implement ${\rm exp}({\rm i}Ht)$.
Consequently, 
%Therefore,
the \QSKSpace will be compiled to a very long gate sequence and constitute the vast majority of cost in post-compilation quantum programs.

Optimizing the compilation of this kernel can immediately benefit a wide range of quantum applications. %while today's compiler optimizations for this important subroutine are far from optimal. 
However, three key challenges have so far hindered deeper compiler optimizations for \QSK s.

%\todo{}

\iffalse
First, the state-of-the-art quantum compilers (e.g., IBM Qiskit~\cite{Qiskit}, Rigetti Quilc~\cite{smith2016practical}) lack a good intermediate representation (IR), like the  control flow graph or static single assignment in classical compilers. 
These quantum compilers always convert input programs to low-level gate sequences before applying optimizations.
The high-level semantics of quantum simulation kernels are lost and hard to be reconstructed from the assembly-style gate sequences. 
Yet, a good IR, which can preserve the high-level algorithmic information, is the key to many high-level  compiler optimizations. 
\fi

First, state-of-the-art quantum compilers (e.g., Qiskit \cite{Qiskit}, Quilc~\cite{smith2020open}, $\rm t|ket\rangle$~\cite{sivarajah2020t}) lack a good formal high-level intermediate representation (IR).
Once programs are converted to low-level gate sequences, the high-level semantics of quantum simulation kernels are lost and hard to reconstruct from assembly-style gate sequences. 
Moreover, simulation kernels face different constraints in different algorithms.
Previous ad-hoc optimizations of quantum simulation~\cite{hastings2015improving,gui2020term,van2020circuit,cowtan2019phase,cowtan2020generic,de2020architecture,vandaele2021phase,shi2019optimized,alam2020circuit,tan2020optimal,li2021software,lao20212qan} are mostly algorithm-specific and do not generalize due to the lack of a formal IR that can uniformly represent simulations kernels as well as varying constraints that are attached to them in different algorithms. %appear in a compiler infrastructure.

Second, most optimizations (e.g., circuit rewriting~\cite{soeken2013white}, gate cancellation~\cite{nam2018automated}, template matching~\cite{maslov2008quantum}, qubit mapping~\cite{murali2019noise}) in today's quantum compilers~\cite{Qiskit, smith2020open}
are local program transformations at small scale. 
However, these passes are designed for generic input program and fail to leverage the deeper optimization opportunities present in quantum simulation kernels which mainly arise from Pauli algebra, flexible synthesis of Pauli strings, and approximations.

%However, the optimization opportunities in compiling quantum simulation kernels come from multiple sources, and to fully leverage them requires a harmonized solution. %, including instruction scheduling, circuit synthesis, gate cancellation, and mapping.
%Existing compiler passes fail to accommodate multiple optimization opportunities collaboratively, leading to limited overall improvement.
%only perform circuit transformations on a very small set of qubits and gates.
%It is hard to find or derive large-scale quantum compiler optimizations because the size of the quantum gate matrices grows exponentially with the number of qubits and a compiler running a classical computer cannot efficiently process these exponentially large matrices.
%Thus, the overall impact of these optimizations is relatively limited especially in a very long gate sequence like the \QSKSpace and a large optimization space is missing. \YS{not sure about this point. compiler optimizations don't use matrices directly.}

%\YS{Compiling simulation kernel is a very hard problem. The optimization opportunities can come from many sources: Pauli algebra, approximation, local structure of gate set. Current compilers cannot capture all these opportunities (local minima), so we need specific algorithm target to handle all these opportunities. (harmonize/reconcile)
%}

Third, quantum simulation kernels appear in a wide range of algorithms. 
Some algorithms~\cite{lloyd1996universal, abrams1999quantum, harrow2009quantum, lloyd2014quantum, rebentrost2014quantum} are designed for fault-tolerant quantum computers with quantum error correction while others~\cite{farhi2014quantum,peruzzo2014variational} target near-term noisy quantum computers. 
The hardware models of these backends can be very different and one single optimization pass may not be suitable for all of them.
%Yet, the optimization opportunities of quantum simulation kernels are mostly from the theory side in a hardware agnostic manner.
Adapting the high-level algorithmic optimizations to the various (and ever-evolving) hardware platforms with different constraints and optimization objectives naturally invokes a reconfigurable compiler infrastructure. %\todo{}
%However, current compilers are mostly monolithic and do not separate general-purpose optimizations and device-specific optimizations well.
%general purpose compilers do poorly and 
%General-purpose compilers do poorly by missing critical device information while device-specific compilers cannot be reused.
%Previous optimizations~\cite{hastings2015improving,gui2020term,van2020circuit,cowtan2019phase,cowtan2020generic,de2020architecture,vandaele2021phase,shi2019optimized,alam2020circuit,tan2020optimal,li2021software} either do not take hardware constraints into consideration or are customized to a specific backend.
%A reconfigurable compiler infrastructure is naturally invoked to support various (and ever-evolving) hardware platforms. %\todo{}
%\YS{"one single compiler optimization algorithm may not be suitable for all of them" -> something like: current compilers do not separate general purpose optimizations + device specific optimizations well , general purpose compilers do poorly and device-specific compilers cannot be re-used.}
%The different characteristics of these various quantum hardware should be considered when 

\begin{figure}[t]
    \centering
    \includegraphics[width=1.0\columnwidth]{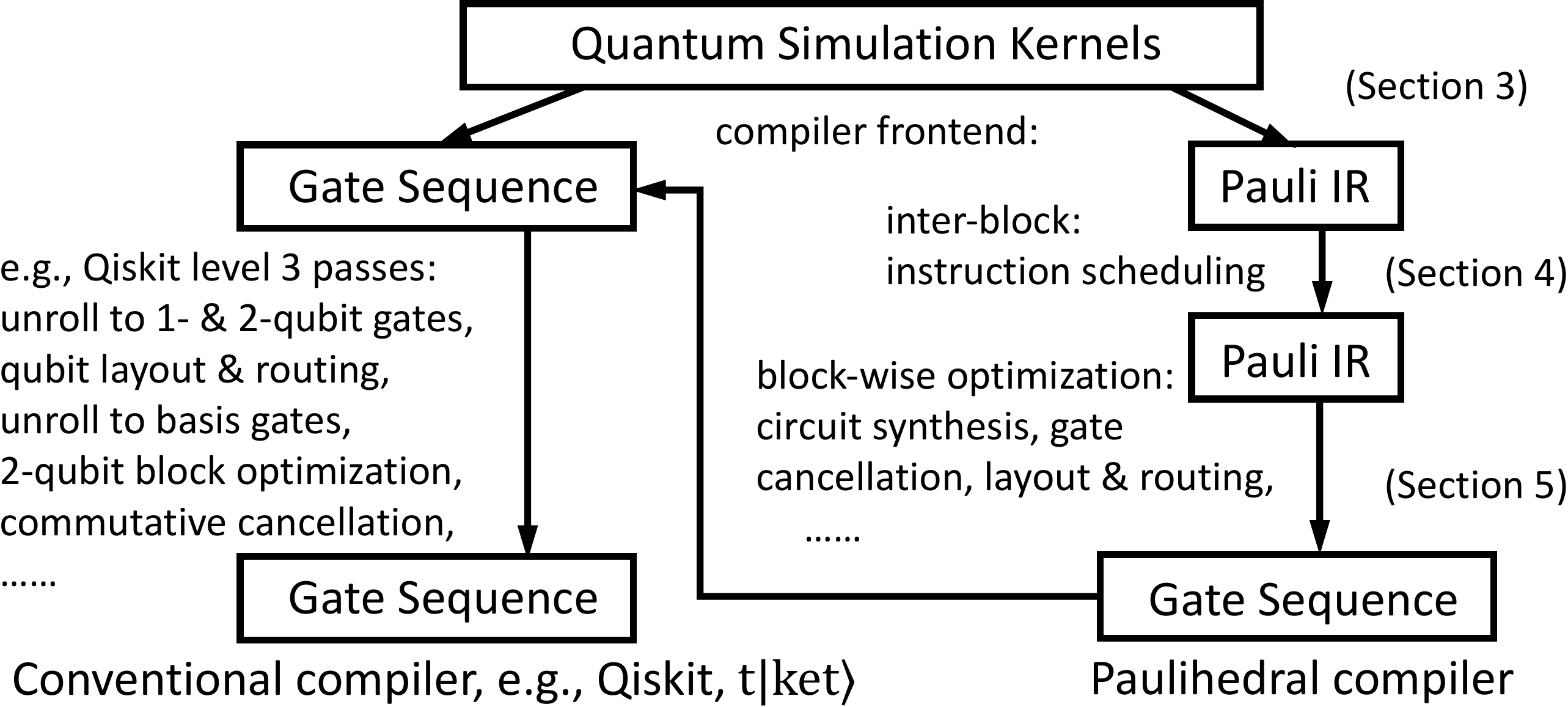}
    \vspace{-15pt}
    \caption{\myCompilerNameSpace vs conventional compilers}%\AJ{typo: kennels --- also PH flow should plug back into Qiskit flow for low-level opts (like shown in results PH+Level3) to show it can work with existing compilers for generic optimization and is not a full replacement.}}
    \vspace{-5pt}
    \label{fig:overallcomparison}
\end{figure}

To overcome these challenges,
we propose \myCompilerName, a compiler framework backed by a formal IR to deeply optimize quantum simulation kernels.
A brief comparison between \myCompilerNameSpace and conventional quantum compilers is shown in Figure~\ref{fig:overallcomparison}.
%\color{blue}
\textbf{First}, \myCompilerNameSpace comes with a new IR, namely Pauli IR, to represent the \QSK s at the Pauli string level rather than the gate level.
The syntax of Pauli IR has a novel block structure which can uniformly represent the simulation kernels of different forms and constraints.
The semantics of Pauli IR is defined on the commutative matrix addition operation. 
Such semantics guarantees that the follow-up high-level algorithmic optimizations are always semantics-preserving and can be safely applied.
%Pauli strings are the basic building blocks of \QSK s and the high-level algorithmic information is maintained.
%\color{black}
\iffalse
\textbf{First}, \myCompilerNameSpace comes with a new IR, namely Pauli IR, to represent the \QSK s at the Pauli string level rather than the gate level.
Pauli strings are the basic building blocks of \QSK s and the high-level algorithmic information is maintained.
The unique algorithmic optimization opportunities are then exposed to the compiler through the new IR, enabling all follow-up optimizations that are hard to be directly implemented in the low-level gate representation.
\fi
%By turning the conventional matrix-multiplication (not commutative in general) based semantics into matrix-addition (always commutative) in the Pauli IR semantics, 
%\myCompilerNameSpace can safely and easily reschedule the instructions throughout the entire program in Pauli IR without changing the high-level program semantics.
%the semantics of a Pauli IR program is defined by the weighted sum of the Pauli strings in all Pauli IR instructions. 
%The addition operation is always commutative, which makes instruction scheduling much more flexible, and \myCompilerNameSpace can easily rewrite the entire program in Pauli IR without changing the high-level program semantics.
\textbf{Second}, we propose several novel optimization passes to reconcile instruction scheduling, circuit synthesis, gate cancellation, and qubit layout/routing at the Pauli IR level.
All these passes are much more effective than their counterparts in conventional gate-based compilers because 
they are operating in a large scope where the algorithmic properties of Pauli strings (\QSK ) are fully exploited.
The optimization algorithms in these passes are also highly scalable since analyzing and processing Pauli strings %, each of which is just an array of $n$ single-qubit operators on the $n$ qubits ($O(n)$ size), 
are much easier than handling the gate matrices on a classical computer.
\textbf{Third}, we decouple the technology-independent and technology-dependent optimizations at different stages and \myCompilerNameSpace can be extended to different backends by adding/modifying the technology-dependent passes. % can be customized to support different hardware backends.
To showcase, we develop technology-dependent optimizations for two different backends, the fault-tolerant quantum computer and the noisy near-term superconducting quantum processor. 
%\YS{though mentioned , I think it should be emphasized her that PauliHedra supports all technologies, e.g ion trap, quantum dots.}

%We expect that \myCompilerNameSpace can benefit many quantum programs because quantum simulation is a widely-used and long-living algorithm design principle.
%\todo{}We perform a comprehensive evaluation for \myCompilerNameSpace over a wide range of \QSK s from various algorithms on both the fault-tolerant and superconducting backends.
%Results show that our depth-oriented scheduling pass can yield  circuits with low depth by trading in a small portion of gate count compared with another gate-count-oriented scheduling pass. And the block-wise optimization passes outperform the baseline (Qiskit at level 3) with both lower gate count and circuit depth.
%We also have real system experiments to show that \myCompilerNameSpace can significantly increase the end-to-end successful rate of QAOA programs on IBM's superconducting quantum devices against the baseline.

Our comprehensive evaluations show that \myCompilerNameSpace outperforms state-of-the-art baseline compilers (Qiskit~\cite{Qiskit}, $\rm t|ket\rangle$~\cite{sivarajah2020t} and algorithm-specific compilers~\cite{alam2020circuit,alam2020efficient,alam2020noise}) with significant gate count and circuit depth reduction on both fault-tolerant and superconducting backends, and only introduces very small additional compilation time.
We also perform real-system experiments to show that \myCompilerNameSpace can significantly increase the end-to-end success rate of QAOA programs on IBM's superconducting quantum devices.

Our major contributions can be summarized as follows:
\begin{enumerate}
%\vspace{-5pt}
    \item We propose \myCompilerName, an extensible algorithmic compiler framework that can deeply optimize \QSK s and thus benefit the compilation of a wide range of quantum programs, with passes that make it retargettable to various backends and optimization objectives.
    %\vspace{-5pt}
  %  \color{blue}
   \item We define a new Pauli IR with formal syntax and semantics which can uniformly represent \QSK s and encode algorithmic constraints of seemingly very different algorithms, and safely expose high-level information to the compiler for optimizations.
    %\item We formally define a new Pauli IR to represent \QSK s  and expose high-level algorithmic information to the compiler for further optimizations.
    %\vspace{-5pt}
    \item We propose several compiler passes for different optimization objectives and backends. 
    They can outperform previous works by systematically leveraging the algorithmic information and they are scalable to efficiently handle larger-size programs.
 %   \color{black}
    %\item We propose several optimization passes in \myCompilerNameSpace  for different optimization objectives and backends. 
    %They can outperform conventional gate-based compilers by leveraging the algorithmic information and they are scalable to efficiently handle larger-size programs.
    %\vspace{-5pt}
    \item Our experiments on 31 different benchmarks show that \myCompilerNameSpace can outperform state-of-the-art baseline compilers with significant gate count and circuit depth reduction. For example, compared with $\rm t|ket\rangle$~\cite{sivarajah2020t},
    %with specialized Pauli string optimizations,
    \myCompilerNameSpace achieves 53.1\% gate count reduction and 53.3\% circuit depth reduction on average on the superconducting backend, as well as 33.6\% gate count and  65.0\% circuit depth reduction on the fault-tolerant backend, using only $\sim 5\%$ additional compilation time.
    For QAOA on a real quantum device, \myCompilerNameSpace achieves end-to-end $1.24\times$ success probability improvement on average (up to $1.87\times$) against the baseline Qiskit compiler~\cite{Qiskit}.
\end{enumerate}

%We have confirmed that the \myCompilerNameSpace compiler infrastructure is being integrated into two industry compilers and we plan to make \myCompilerNameSpace open source.
%We expect that \myCompilerNameSpace can benefit many quantum programs because quantum simulation is a widely-used and long-living algorithm design principle.

\section{Background}\label{sec:background}
In the section we introduce the necessary background about quantum simulation kernels.
We do not cover basic quantum computing concepts (e.g., qubit, gate, linear operator, circuit) and we recommend~\cite{nielsen2010quantum} for more details.

\subsection{Pauli String and Compilation}
We start with the Pauli string, the basic concept
%building block
in quantum simulation.
For an $n$-qubit system, a \psSpace is defined as $P = \sigma_{n-1} \sigma_{n-2}\cdots \sigma_{0}$ where $\sigma_i \in \{I, X, Y, Z\}$, $0\leq i \leq n-1$. $X$, $Y$, $Z$ are the three Pauli operators, and $I$ is the identity. % operator. 
$\sigma_i$ corresponds to the $i$-th qubit. % in the $n$-qubit system.
%One \psSpace $P$ naturally
The operators in a Pauli string $P$ can represent a Hermitian operator $\otimes_{i=0}^{n-1}\sigma_i$ ($\otimes$ is the Kronecker product), which can be denoted by $P$ without ambiguity.
In the rest of this paper, we do not distinguish a Pauli string $P$ and the Hermitian operator generated by $P$.

\begin{figure}[b]
    \centering
    \vspace{-1pt}
    \includegraphics[width=1.0\columnwidth]{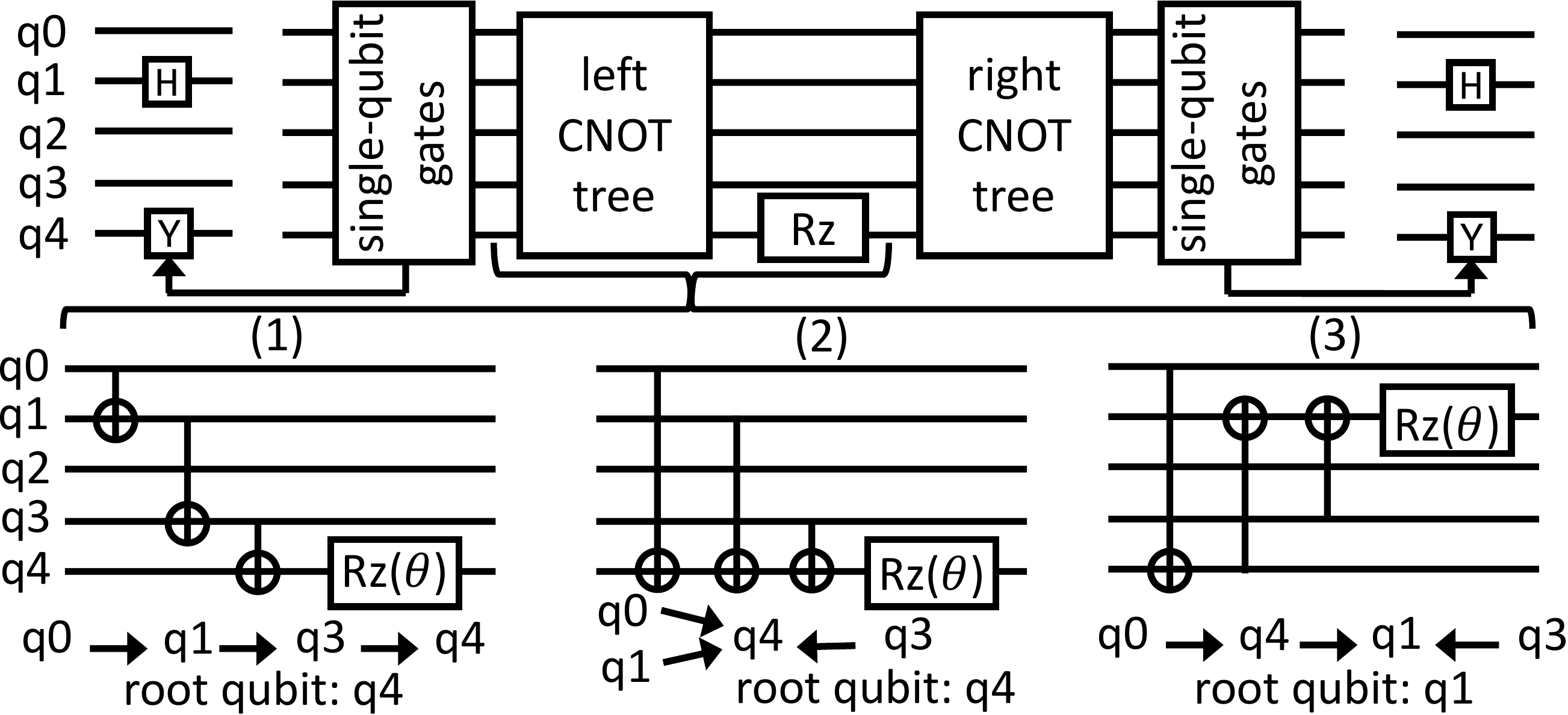}
    \vspace{-17pt}
    \caption{Synthesis example of ${\rm exp}({\rm i}Y_4Z_3I_2X_1Z_0\frac{\theta}{2})$}
    \vspace{-2pt}
    \label{fig:PSsynthesis}
\end{figure}

One important property of a \psSpace is that the operator ${\rm exp}({\rm i}P\frac{\theta}{2})$ can be easily synthesized into basic gates.
%as shown in in Figure~\ref{fig:PSsynthesis}.
%Suppose we wish to synthesize ${\rm exp}({\rm i}Y_4Z_3I_2X_1Z_0\frac{\theta}{2})$.
Figure~\ref{fig:PSsynthesis} shows an example of synthesizing ${\rm exp}({\rm i}Y_4Z_3I_2X_1Z_0\frac{\theta}{2})$.
%The synthesis result is on the top of Figure~\ref{fig:PSsynthesis}.
There are two identical layers of single-qubit gates at the beginning and the end of the synthesized circuit.
In this single-qubit gate layer, there are $\rm H$ or $\rm Y$ gates on those qubits whose operators are $X$ (i.e., $q1$) or $Y$ (i.e., $q4$) in the Pauli string, respectively.
%In this example, we have $\rm H$  gate on $q1$ and $\rm Y$  gate $q4$. % in the single-qubit gate layers.
In the middle is a left $\rm CNOT$ tree, a central $\rm Rz(\theta)$ gate, and a right  $\rm CNOT$ tree.
The left % $\rm CNOT$ 
tree can be generated in different ways and the only requirement is to connect all the qubits whose operators are not the identity in $P$ (e.g., $q0$, $q1$, $q3$, $q4$ in Figure~\ref{fig:PSsynthesis}).
%In this example, we need to connect $q0$, $q1$, $q3$, $q4$.
The lower half of Figure~\ref{fig:PSsynthesis} shows three different but valid ways to generate the $\rm CNOT$ tree circuits and their corresponding tree graphs. 
In these trees, the $\rm CNOT$ gates should connect the qubits from the leaf nodes to the root node.
Any qubit in the tree can become the root (e.g.,  $q4$  in Figure~\ref{fig:PSsynthesis} (1) (2), $q1$ in Figure~\ref{fig:PSsynthesis} (3)). % node qubit.
%For example, $q4$ is the root node qubit in the first two implementations while $q1$ is the root in the third implementation.
The central $\rm Rz(\theta)$ gate is applied on the root qubit and the right $\rm CNOT$ tree has the same $\rm CNOT$ gates in the left tree but in a reversed order. Paulihedral uses this algorithmic flexibility in synthesis to increase gate cancellation and also reduce mapping overhead.

%\YS{mention commutation relations between X Y Z and strings? This helps people understand some optimizations}

\subsection{Quantum Simulation Kernels}
\begin{figure}[t]
    \centering
    \includegraphics[width=1.0\columnwidth]{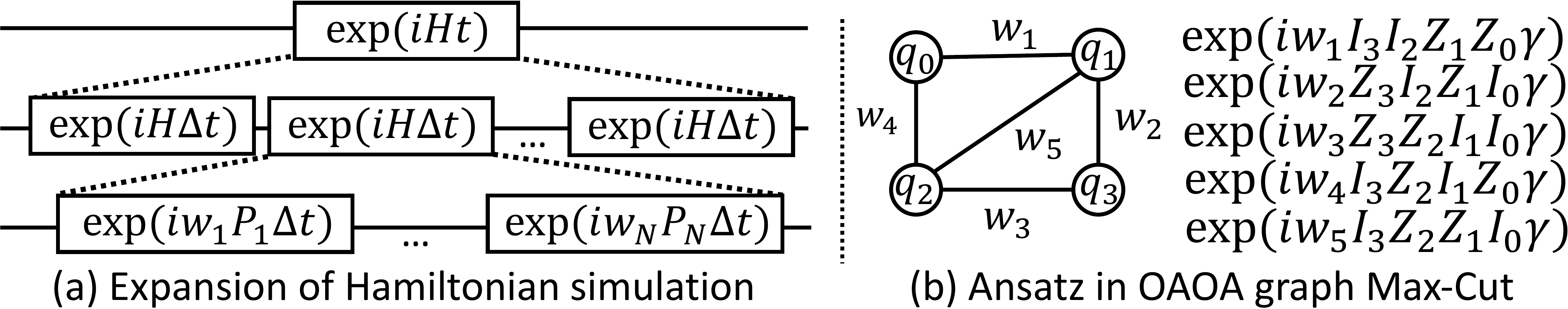}
    \vspace{-15pt}
    \caption{Example of \QSK s}
    \vspace{-5pt}
    \label{fig:kernelexample}
\end{figure}

The \QSKSpace is to (approximately) implement the operator ${\rm exp}({\rm i}Ht)$ where $H$ is the Hamiltonian of the simulated system and $t \in \mathbb{R}$.
Since directly compiling ${\rm exp}({\rm i}Ht)$ into single- and two-qubit gates is hard, a compiler usually expands $H$ in the Pauli basis, i.e., $H=\sum_{j=1}^{N}w_jP_j$ where $w_j \in \mathbb{R}$ and $P_j$ is a Pauli string.
Then ${\rm exp}({\rm i}Ht)$ is approximated using the Trotter formula~\cite{trotter1959product}:
${\rm exp}({\rm i}Ht) = \left[\prod_{j=1}^{N}{\rm exp}({\rm i}P_jw_j\Delta t)\right]^{\frac{t}{\Delta t}} + O(t\Delta t)$.
$\Delta t$ is a parameter determined by the simulation accuracy.
Figure~\ref{fig:kernelexample} (a) shows the expansion process. ${\rm exp}({\rm i}Ht)$ is first converted to %the multiplication of
$\frac{t}{\Delta t}$ terms of ${\rm exp}({\rm i}H\Delta t)$. Each ${\rm exp}({\rm i}H\Delta t)$ is then expanded to an array of ${\rm exp}({\rm i}P_jw_j\Delta t)$ 
and converted to basic gates.
%Finally, each term ${\rm exp}({\rm i}P_jw_j\Delta t)$ can be compiled to basic gates as mentioned above.

\iffalse
Quantum simulation kernels also appear in the recently developed promising variational quantum algorithms. 
In these algorithms, the vast majority of the program is the ansatz, which is a quantum circuit with tunable parameters to guess an answer to the target problem.
One popular type of ansatz is the application-inspired ansatz which usually has good trainability and is more likely to converge to a better solution~\cite{cerezo2020variational}.
The design of application-inspired ansatz is inspired by the quantum simulation principle and it can be considered as implementing a set of operators which are the simulation of the Pauli strings from the Hamiltonian of the target problem.
\fi
%In the recently developed promising variational quantum algorithms, the quantum simulation kernels also appear as the an
Quantum simulation kernels also appear in recently developed variational quantum algorithms, in which the vast majority of the program is an ansatz (parameterized quantum circuit). %, also known as ansatz. 
One popular type of ansatz with good trainability is the application-inspired ansatz~\cite{cerezo2020variational} which can be considered as a simulation kernel.
Compared with implementing ${\rm exp}({\rm i}H\Delta t)$, the only difference is that the $\Delta t$ is changed to some tunable parameters associated with different Pauli strings and \textit{the overall program structure remains the same}.
For example, Figure~\ref{fig:kernelexample} (b) shows the ansatz of QAOA algorithm~\cite{farhi2014quantum} on a 4-node graph Max-Cut problem.
The graph of the problem has 5 edges of different weights, and the Hamiltonian of this problem is the weighted sum of the 5 Pauli strings associated with the 5 edges.
The majority of the QAOA ansatz~\cite{farhi2014quantum} is to implement the 5 operators on the right ($\gamma$ is the parameter). % so it can also be considered as a \QSK.

%(e.g., UCCSD~\cite{peruzzo2014variational} for chemistry simulation)
%These ansatzes usually have good trainability and are more likely to converge to a better solution.

%\subsection{Simulation Kernels in Variational Quantum Algorithms}

\section{Foundations of Paulihedral}\label{sec:overview}
%To express the \QSK s at a high-level and expose the

%In this section, we first introduce the unique optimization opportunities in \QSK s and the challenges to exploit them in conventional compilers. 
%Then we introduce a new Pauli-string-based IR, which is the foundation of all follow-up optimization passes in \myCompilerName.

%In this section, we introduce the foundations of \myCompilerName. 
In this section, we first introduce the opportunities and challenges of compiler optimizations for the simulation kernel.
%we 
%We 
%start from the unique optimization opportunities in \QSK s and the challenges to exploit them in conventional compilers. 
Then we formally introduce a new IR that maintains the high-level information and the algorithm constraints in \myCompilerName.% enables all follow-up optimization passes in \myCompilerName.

\begin{figure}[t]
    \centering
    \includegraphics[width=0.95\columnwidth]{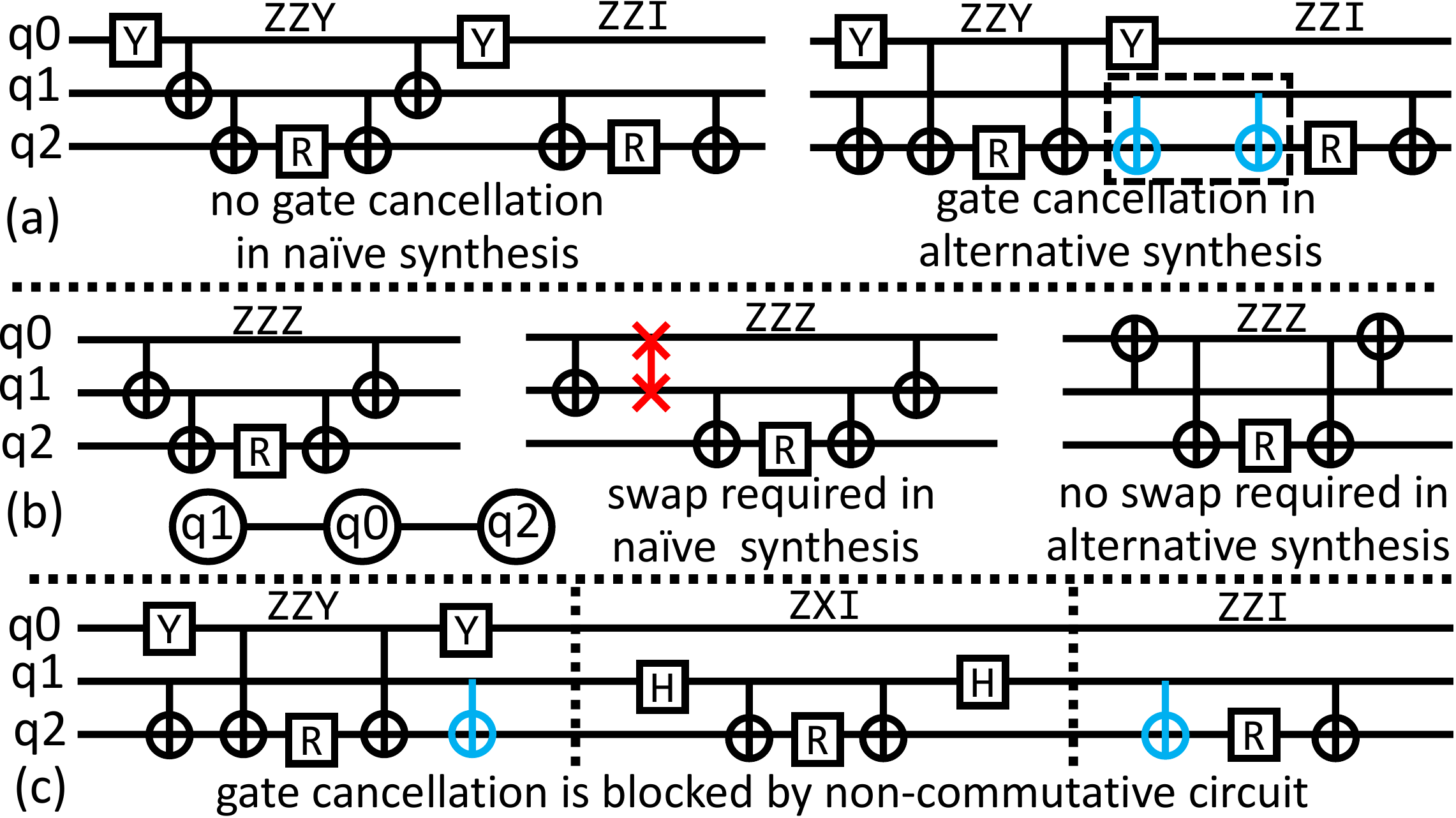}
    \vspace{-5pt}
    \caption{Optimization opportunities and challenges%\AJ{make endianness consistent with figure 2: should be ZZY, ZZI, ZXX. in this figure.}
    } %\AJ{in (b) I think there should be a gap in the wires between the 1st and 2nd ZZZ. in (c) shouldn't you use the YZZ synthesis that allows the gate cancellation?}}
    %\vspace{-5pt}
    \label{fig:challenge}
\end{figure}

\subsection{Opportunities and Challenges}
The optimization opportunities used in this paper come from the properties of Pauli strings mentioned above. We introduce them by the examples in Figure~\ref{fig:challenge}. 
\textbf{1) Gate cancellation:} It is possible to have more gate cancellation by selecting a different synthesis plan for the ${\rm exp}({\rm i}P\theta)$. %\AJ{everything after here should be reversed endian ZZY, ZZI, etc.}
Suppose the naive synthesis is the one in Figure~\ref{fig:PSsynthesis} (1) and we have two Pauli strings, $ZZY$ and $ZZI$.
Under the naive synthesis (on the left of Figure~\ref{fig:challenge} (a)) there is no gate cancellation. 
However, in an alternative synthesis of $ZZY$, we can have two $\rm CNOT$ gates cancelled (on the right of Figure~\ref{fig:challenge} (a)).
%In general, for the circuit blocks of two Pauli strings, the more qubits on which the two strings share the same Pauli operator, the more gates can be cancelled.
%\todo{mention in Section 4}
\textbf{2) Mapping:} The mapping overhead onto connectivity-constrained architectures can also be reduced. % by alternative synthesis.
For example, we wish to map the $ZZZ$ simulation circuit onto a linear architecture with the current mapping shown in Figure~\ref{fig:challenge} (b).
Under the naive synthesis we need to insert one SWAP between $q0$ and $q1$.
While a better synthesis plan on the right of Figure~\ref{fig:challenge} (b) does not require any SWAPs.

Although there is much optimization space for \QSK s, such optimizations are not yet widely deployed in today's quantum compiler infrastructures due to the following challenges.
\textbf{1) Missing high-level information:} Once the program is converted to basic gates, where today's compilers perform most optimizations, it is hard to identify and reconstruct the high-level semantics of Pauli string simulation circuit blocks from an assembly-style gate sequence.
%Given a Pauli string, it is easy to synthesize the corresponding circuit, but the opposite direction, i.e. identifying and reconstructing the high-level semantics of Pauli string simulation circuit blocks from an assembly-style gate sequence, is highly non-trivial (e.g., very complex template matching).
%Therefore, it becomes hard to leverage the optimization opportunities mentioned above since the compiler optimization passes do not know which part in a long gate sequence constitutes a simulation circuit.
%\textbf{2) Missing constraint encoding:} The simulation kernels of different algorithm face different constraints during compilation. Some kernels may require implementing some Pauli string simulations together for algorithmic optimization purposes like symmetry preserving~\cite{gard2020efficient}, parameter sharing~\cite{farhi2014quantum,peruzzo2014variational}, error suppression~\cite{gui2020term}, simultaneous diagonalization~\cite{van2020circuit},
\textbf{2) Non-semantics-preserving optimization:} To leverage some optimization opportunities would require non-semantics-preserving operations that are usually not allowed in a compiler.
%Even if the compiler knows the circuit blocks are for Pauli string simulation, some optimization opportunities cannot be fully exploited. 
For example, consider the program in Figure~\ref{fig:challenge} (c).
It is known from Figure~\ref{fig:challenge} (a) that gates can be cancelled between $ZZY$ and $ZZI$ but now there is an $ZXI$ simulation circuit between them.
We observe that the order of the simulation terms with respect to different \ps s is not specified in the Trotter formula or the variational form requirement.
So, from an algorithmic perspective, the compiler may exchange the order of $ZZI$ and $ZXI$, making $ZZY$ and $ZZI$ adjacent for gate cancellation.
%Then $YZZ$ and $IZZ$ are adjacent and we can do the alternative synthesis for gate cancellation.
However, such operation is not semantics preserving from a gate-level perspective because, in general, ${\rm exp}({\rm i}ZZI\theta_1){\rm exp}({\rm i}ZXI\theta_2)\neq{\rm exp}({\rm i}ZXI\theta_2){\rm exp}({\rm i}ZZI\theta_1)$. This would be impossible to leverage without an IR that is able to encode such algorithmic knowledge.
%Conventional compilers usually do not allow such non-semantics-preserving transformations.

% As introduced in the last section, the synthesis of Pauli string simulation circuits can be very flexible and such flexibility bring 

%It can be observed that the order of the simulation terms with respect to different \ps s is not strictly specified in the Trotter formula.
%Once we implement the simulation of all the Pauli strings in the Hamiltonian expansion, the overall approximation is always bounded by the small time step $\Delta t$.

\subsection{Pauli IR: Syntax and Semantics}
To overcome the challenges above, the objective of the new IR is to maintain high-level algorithmic information %encode the application constraints\todo{}, 
and make all transformations semantics-preserving.
Our new IR, namely Pauli IR, realizes them with its syntax and semantics.
\begin{figure}
    \centering
    \vspace{-5pt}
    {\footnotesize
    \begin{align*}
    \langle program \rangle  &::= \langle pauli\_block\rangle \\
    & ~~~ | \ \langle program\rangle \ ; \ \langle pauli\_block\rangle \\
    \langle pauli\_block\rangle & ::= \{\langle pauli\_str\_list\rangle, parameter\} \\
    \langle pauli\_str\_list\rangle & ::= \langle pauli\_str, weight \rangle \\
    & ~~~ | \ \langle pauli\_str\_list\rangle \ ; \ \langle pauli\_str, weight \rangle  \\
    \langle pauli\_str \rangle & ::=  \sigma_{n-1} \sigma_{n-2} \cdots \sigma_0\\
    \sigma_i & ::= I \ | \ X \ | \ Y \ | \ Z, \ (0 \leq i \leq n-1) \\
    parameter,\ weight  & ~\in \mathbb{R} %~ [-2\pi, \ 2\pi] % \mathbb{R}
\end{align*}
} %
\vspace{-17pt}
    \caption{Formal syntax of an $n$-qubit Pauli IR program}
    \vspace{-5pt}
    \label{fig:syntax}
\end{figure}
%\vspace{-2pt}
    
%\end{definition}
%\vspace{-1pt}

\begin{figure}[t]
    \centering
    \includegraphics[width=1.0\columnwidth]{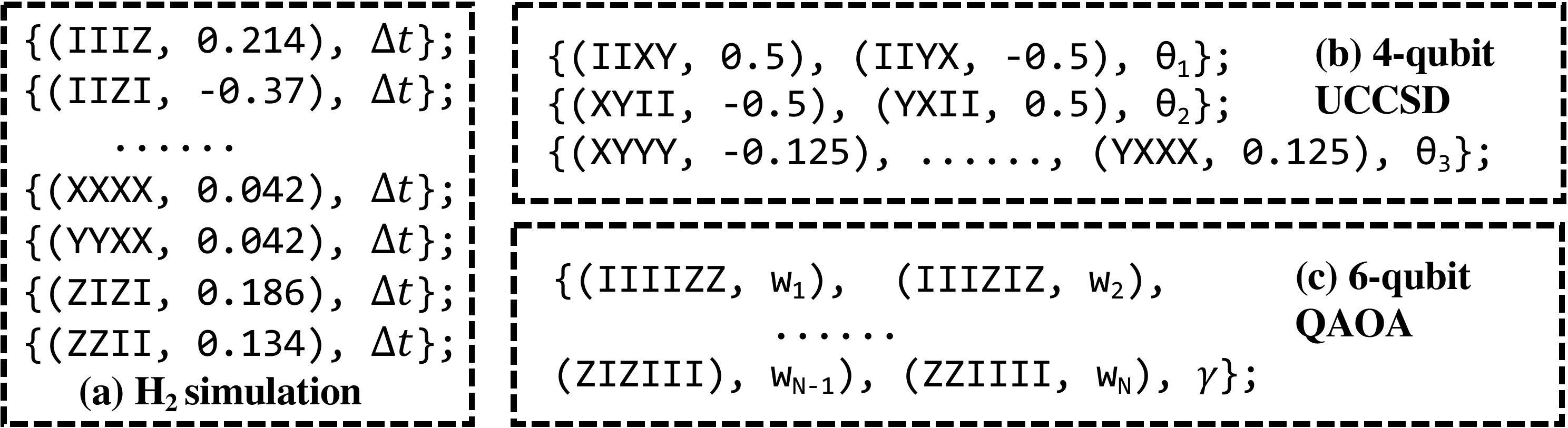}
    \vspace{-15pt}
    \caption{Example Puali IR programs}
    %\vspace{-5pt}
    \label{fig:example-program}
\end{figure}

\textbf{Syntax:} The syntax is shown in Figure~\ref{fig:syntax} and explained as follows.
A $program$ is recursively defined as a list of $pauli\_block$s. Each $pauli\_block$ is a tuple with two elements. 
The first element is a list of weighted Pauli strings ($pauli\_str\_list$s) and the second element is a real-valued $parameter$ shared by all Pauli strings in this $pauli\_block$. 
One element in the $pauli\_str\_list$ is an $n$-qubit Pauli string and a real-value $weight$.
Figure~\ref{fig:example-program} shows the Pauli IR code of three example programs. 
Figure~\ref{fig:example-program} (a) simulates the Hamiltonian of $\rm H_2$ and each $pauli\_block$ has one $pauli\_str$.
Figure~\ref{fig:example-program} (b)(c) are variational quantum algorithms so that parameters are labeled by $\theta$ and $\gamma$.
In the UCCSD program (Figure~\ref{fig:example-program} (b)), each $pauli\_block$ has multiple $pauli\_str$s which share the same $\theta$ in the $pauli\_block$.
In the QAOA program (Figure~\ref{fig:example-program} (c)), all $pauli\_str$s are in one $pauli\_block$, sharing the parameter $\gamma$.
%with the shared parameter $\gamma$.

\textbf{Encoding constraints:}
%It is possible to represent \QSK s with only a list of $\langle pauli\_str, weight\rangle$.
One key advantage of the IR syntax is that the algorithmic constraints in all simulation kernels, as far as we know,  can be naturally encoded.
%The reason that we design the $pauli\_block$ in Pauli IR is that,
In some %\QSK s
simulation kernels
(e.g., UCCSD~\cite{peruzzo2014variational}, QAOA for constrained optimization~\cite{saleem2020approaches}), the algorithm requires that some %(usually mutually commutative)
Pauli strings should always appear together %consecutively 
for some algorithmic purposes like symmetry preserving~\cite{gard2020efficient}, parameter sharing~\cite{farhi2014quantum,peruzzo2014variational}, error suppression~\cite{gui2020term}, etc.
%, simultaneous diagonalization~\cite{van2020circuit}, etc.
Pauli IR employs a $pauli\_block$ structure to represent such constraints.
 The compiler can extract such information and all the $pauli\_str$s inside one $pauli\_block$ are always scheduled together in follow-up optimization passes.
%during the the instruction scheduling in \myCompilerNameSpace to avoid losing these algorithm-level optimizations.
In the rest of this paper, $pauli\_block$ is denoted by block for simplicity.
%$pauli\_block$ is denoted as block for simplicity in the rest of this paper.
%\myCompilerNameSpace only performs free scheduling at the level of $pauli\_block$s and 
%This further automates other algorithm optimizations in \myCompilerName.

\textbf{Semantics:}
The IR's semantics function (denoted by $\sem{\langle program\rangle}$) can be formally defined by the rules in Figure~\ref{fig:semantics}. 
This function is a mapping from the IR syntax to the set of all Hermition operators in a $2^n$-dimensional Hilbert space as our IR is to represent the Hamiltonian to be simulated.
Note that the rules in the second and the fourth rows are defined based on \textit{matrix addition} which is always commutative.
As a result, exchanging the order of the $pauli\_block$s in a $program$ or the order of $\langle pauli\_str, weight \rangle $s in a $pauli\_block$ will not change the semantics.

\begin{figure}[t]
    \centering
    \vspace{-5pt}
    { \footnotesize
    \begin{align*} 
    \sem{\varnothing}= & \ 0 \\ 
    \sem{\langle program\rangle;\langle pauli\_block\rangle}= & \ \sem{\langle program\rangle\rrbracket + \llbracket \langle pauli\_block\rangle} \\
     \sem{\{\langle pauli\_str\_list\rangle, parameter\} }=  & \ parameter \times \sem{\langle pauli\_str\_list\rangle} \\
     \sem{\langle pauli\_str\_list\rangle;\langle pauli\_str, weight \rangle}= & \  \sem{\langle pauli\_str\_list\rangle} \\
    &+ \sem{\langle pauli\_str, weight \rangle}\\
      \sem{\langle pauli\_str, weight \rangle }= &  \ weight \times \sem{pauli\_str} \\
   \llbracket \sigma_{n-1} \sigma_{n-2} \cdots \sigma_0 \rrbracket=  & \ \sigma_{n-1} \otimes \sigma_{n-2} \otimes\cdots\otimes \sigma_0
\end{align*}
 }%
\vspace{-17pt}
    \caption{Formal semantics of an $n$-qubit Pauli IR program}
    \vspace{-5pt}
    \label{fig:semantics}
\end{figure}

\begin{figure*}[t]
    \centering
    \includegraphics[width=0.95\textwidth]{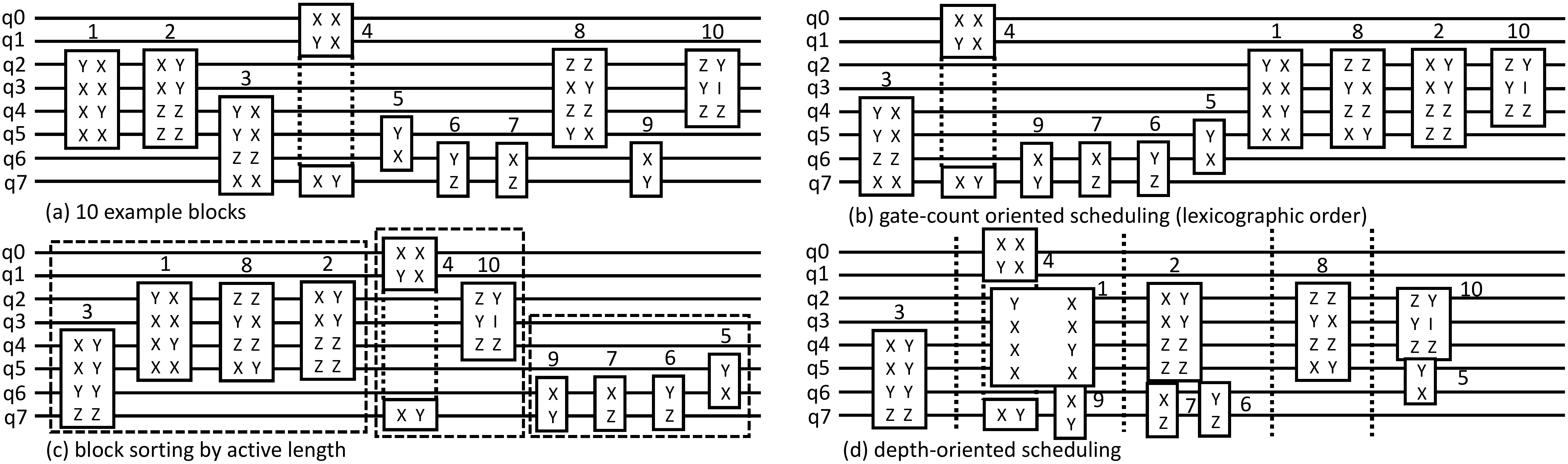}
    \vspace{-5pt}
    \caption{Example of block scheduling optimizations}
    %\ANBANG{(d) block 8 and block 2 order is not right}}
    %\vspace{-5pt}
    \label{fig:blockscheduling}
\end{figure*}

\section{Block-Wise Instruction Scheduling Passes}\label{sec:interblock}

The first step in \myCompilerNameSpace
%compiling a Pauli IR program 
is to schedule the blocks and the instructions within each block.
%using the inter-block optimization in \myCompilerName. 
%As discussed in Section~\ref{sec:overview}, 
Intuitively for two adjacent Pauli strings, more gates can be cancelled if they share the same non-identity operators on more qubits. % \todo{}
Trying to maximize the number of shared operators between consecutive strings would be desirable. % can result in a small number of gates in the post-compilation circuit.
Also, it is possible to execute multiple blocks which have non-identity operators on disjoint sets of qubits in parallel and reduce the final circuit depth. 
In this paper, we present two block scheduling algorithms for two optimization objectives, reducing the total gate count or the circuit depth.
%In this paper, we implement the lexicographic order as a gate-count-oriented scheduling and propose a new depth-oriented scheduling algorithm.
We explain our block scheduling optimizations using the example in Figure~\ref{fig:blockscheduling}. %with %.
Suppose %that 
we will schedule 
10 blocks on 8 qubits (Figure~\ref{fig:blockscheduling} (a)).
In these blocks, each column is a Pauli string. 
The identity operators are omitted since they do not result in any circuit.

\subsection{Gate-Count-Oriented Scheduling}
Lexicographic ordering of Pauli strings has been shown to be effective at enabling gate cancellation between them~\cite{Tranter2018comparison,gui2020term}. Here we adapt this principle to the multi-string-per-block case for our gate-count-oriented scheduling algorithm.
In the lexicographic order, the Pauli strings are scheduled in the alphabetical order.
In Figure~\ref{fig:blockscheduling}, we assume $X < Y < Z < I$ and use little-endian to lexicographically order from $q7$ down to $q0$.
%\AJ{also I think need to say ordering is by qubits and in Fig7 we use little-endian to lexicographically order from q7 down to q0.}
%We extend the lexicographic order to the multi-string-per-block cases.
When a block has multiple strings, we first apply the lexicographic order on all %the Pauli 
strings in %side
this block and then use the first %Pauli 
string to represent this block when compared with other blocks.
The first Pauli string can be representative because the strings in one block are usually mutually commutative in practical %Pauli IR 
algorithmic constraints~\cite{peruzzo2014variational,farhi2014quantum,gui2020term}.
%(e.g., UCCSD~\cite{peruzzo2014variational}). 
Two strings in a mutually commutative set can share the same operators on many qubits and all strings in one block are similar.
Figure~\ref{fig:blockscheduling} (b) shows the result of gate-count-oriented scheduling. % (assuming $X < Y < Z < I$). % \todo{}
%It can be observed that lexicographic order can create many gate cancellation opportunities.
%For example, block 3 and 4 share $X$ on $q7$. Block 2 and 10 share $Z$  on $q4$ and $Y$ on $q3$.
%\AJ{if you choose the 1st Pauli string as representative of the whole block, then these common operators seem accidental? i.e. $Y$ is common b/w block 2 and 10 but the algorithm only looked at the 1st column, which does not contain $Y$ for block2.}
%\todo{we extend}

%\setlength{\textfloatsep}{0.1cm}
%\setlength{\floatsep}{0.1cm}

\setlength{\textfloatsep}{8pt}
\setlength{\floatsep}{0pt}

\begin{algorithm}[t]
%\setstretch{1}
\SetAlgoLined
 
\KwIn{List of Pauli blocks.}
\KwOut{Pauli Layers $L$.}
Sort Pauli blocks by active-length-decreasing order, then sort blocks of the same active length by lexicographic order\;

 $R$ = the set of all Pauli blocks remaining; %\tcp{R is the set of Pauli blocks that has not been placed in a layer yet.}
 $L = \varnothing$\; % \tcp{L is the set of constructed Pauli layers consisting of Pauli blocks.}
 Initialize the first layer\;
 \While{$R$ is not empty}{
 
% \tcc{Find one block for a new layer}
 $next\_block = \argmax_{block\in R} \rm Overlap(block, \text{last Pauli layer})$\; 
 %\tcp{the maximum overlap between the last Pauli layer and a Pauli block}
% $next\_pb = NULL$\;
% $min\_dist = 0$\; 

% \For{\textbf{pb} in R}{
%    \If{hamming_dist(\textbf{pb}, last Pauli layer) $< min\_dist$ }{
%        $next\_pb = \textbf{pb}$\;
%        $min\_dist = hamming\_dist(\textbf{pb}, last\ Pauli\ layer)$\;
%    }
% }
 $pauli\_layer = [next\_block]$; $R.remove(next\_block)$\;
 %\tcp{Adding new Pauli layer}
 %$R.remove(next\_pb)$\;
% \tcc{padding the new Pauli layer}
 \While{total depth of the small padding blocks < the depth of $next\_block$}{
 find small Pauli block not overlapped with $next\_block$\;
  Append these blocks to $pauli\_layer$\;
    Remove these blocks from $R$\;
 }
 
 \iffalse
 \Repeat{$pauli\_layer$ converges}{
    \Repeat{total depth of these blocks exceeds the depth of $next\_block$}{find small Pauli block not overlapped with $next\_block$\;}
    Append these blocks to $pauli\_layer$\;
    Remove these blocks from $R$\;
 }
 \fi
$L.append(pauli\_layer)$\;
}

%\v%space{-10pt}
\caption{Depth-oriented scheduling}
\label{alg:DOscheduling}
%\vspace*{-10pt}

\end{algorithm}

\subsection{Depth-Oriented Scheduling}
%The lexicographic order is good at cancelling gates while it may result in a long depth circuit. 
The blocks can also be scheduled for reducing circuit depth.
For example, in Figure~\ref{fig:blockscheduling}, $q0$ to $q5$ are idle when executing block 9, 7, and 6. 
We may execute block 1 with them in parallel so that the overall circuit depth
%of the generated circuit
can be %further
reduced.
We propose a new depth-oriented block scheduling algorithm, whose pseudocode is in Algorithm~\ref{alg:DOscheduling}.
%We explain our algorithm using the same example in Figure~\ref{fig:blockscheduling}.
For the example in Figure~\ref{fig:blockscheduling},
we first sort all blocks by the active length of the Pauli strings of the blocks in a decreasing order. 
%We first pre-process all blocks by ordering them by the active length of the Pauli strings of the blocks, in decreasing order. 
%The effective length is defined to be the number of non-identity operators in a \ps since its simulation circuit will have gates only on those qubits with non-identity operators. 
%For the $pauli\_block$s with multiple Pauli strings, its
The active length of a  block is defined by the number of qubits which have a non-identity operator in at least one Pauli string of this block.
%If there is only one string in this block, the active length is the number of non-identity operators in this string.
This is an over-approximated estimation on how a block will occupy the qubits.
%For the blocks of the same active length, we order them by the lexicographic order above.
The blocks of the same active length are ordered by the lexicographic order above.
Figure~\ref{fig:blockscheduling} (c) shows the sorting result.
Block 3, 1, 8, 2 have the largest active length of 4 so they are at the beginning.
Block 9, 7, 6, 5 have the smallest active length of 2 and they are at the end.
%\vspace{-5pt}

Then we begin to schedule all blocks and put the blocks in different layers to increase the parallelism. 
%The key idea is organize blocks in layers. 
For each layer, we first schedule a large active length block. 
Then we search for small active length blocks that can be executed in parallel with the large block.
For the example in Figure~\ref{fig:blockscheduling} (d), we initialize the first layer by selecting the first block after the sorting.
%This block has the largest active length.
%For example (Figure~\ref{}), 
We place the block 3 at the beginning.
Then we search for  small  blocks that have no overlapped active qubits with the large block and can be executed in parallel.
%can be executed in parallel with this large block. 
%These small blocks must have no overlapped active qubits with the large block.
There are no such small blocks for block 3 
%\AJ{why doesn't block 4 work?} 
so we continue by start another layer with block 1.
In this layer, block 4, 9, 7, 6 can be placed in parallel with block 1.
We iterate over the sorted block and find the first few blocks that can padded in this layer. 
In this example, we select block 4 and 9. 
%We repeat this process to find a series of $pauli\_block$s.
We also estimate the depth of these small blocks so that total depth of these  blocks will not exceed the depth of the original large block in this layer.
%After this round of padding, we will update the active qubits by adding the active qubits in the series of padded small blocks.
We repeat this padding process until we cannot find any new blocks that can be added in this layer.
%For example. in Figure~\ref{}, the first layer is.
We then continue to the next layer and start with block 2 because its first Pauli string has the most overlapped Pauli operators with the Pauli strings at the end of the previous layer.
%next large block (block 8).
% in Figure~\ref{fig:blockscheduling}).
%(block 8) that has the most overlapped operators with the  Pauli strings at the end of the previous layer.
%in Figure~\ref{}, we select block ?? in the second layer.
We iterate until all  blocks are scheduled.
Figure~\ref{fig:blockscheduling} (d) shows the final  result of our depth-oriented scheduling and we can expect that the circuit depth can be reduced even if we do not convert the program to the gates. 
%Although we are still at the block level and have not reached the actual gate-level implementation, we can expect that the circuit depth can be reduced with this depth-oriented scheduling algorithm.
This is another benefit of Pauli IR because the compiler can operate on a fairly compact description of the program. 
Once the program is lowered to gates, then the size blows up and parallelizing gates becomes much more expensive.

%\AJ{How expensive is this? can we say something like ``this is another benefit of Pauli IR because the compiler can operate on a fairly compact description of the program. Once the program is lowered to gates, then the size blows up and parallelizing gates becomes much more expensive''}

%Besides this depth-oriented (DO) block scheduling, we also implemented lexicographic (LEX) order which 
%\setlength{\textfloatsep}{\textfloatsepsave}
%\setlength{\floatsep}{\floatsepsave}

\section{Block-Wise Optimization Passes}\label{sec:innerblock}

%After the blocks are scheduled, we need to convert the Pauli IR program into a gate sequence to execute the program.
%to be executable.
%As discussed in Section~\ref{sec:background} and~\ref{sec:overview}, the synthesis of Pauli string simulation circuit has great flexibility.
In this section, we introduce two optimization passes that can exploit the gate cancellation potential created by our scheduling passes in the last section, and convert the Pauli IR programs to gate sequences with different optimization objectives onto the fault-tolerant quantum computer (FT) backend and the near-term superconducting quantum computer (SC) backend.
\iffalse
The scheduling passes in  Section~\ref{sec:interblock} are designed to increase the gate cancellation potential by increasing the overlapped operators between two consecutive Pauli strings.
However, careful optimization when converting to gate sequences is still required to fully realize this potential.
%In this section, we will synthesize each Pauli string in the Pauli IR program into a gate sequence.
%The tree structure of the CNOT gates and the root qubit will be exploited to realize as much gate cancellation as possible.
Since different hardware backends have different properties and constraints, \myCompilerNameSpace employs different 
%inner-block\AJ{intra-block?}
block-wise optimization passes to generate gate sequence from Pauli IR program with different optimization objectives.
In this section, we introduce two passes onto the fault-tolerant quantum computer (FT) and the near-term superconducting quantum computer (SC).

\fi

\subsection{On the Fault-Tolerant Backend}\label{sec:BCforFTQC}

%On the fault-tolerant (FT) backend, the cost is dominated by the gate count (specifically T gates) and  the mapping overhead can be neglected  after applying the quantum error correction~\cite{fowler2012surface}. %\todo{}
%The common optimization objective is to reduce the gate count and circuit depth~\cite{van2020circuit,hastings2015improving,cowtan2019phase,cowtan2020generic}.
%Therefore, the pass for compiling on the FT backend will try 

%We assume there is no mapping requirement (logical CNOT is directly supported on two arbitrary logical qubits).\AJ{I think instead of saying CNOT is directly supported between any 2 qubits (in reality there is braiding etc.), we should say the cost is dominated by gate count (specifically T gates) and so we neglect the mapping overhead cost.}
%All logical qubits are uniform and there is no significant variance among the gates on different logical qubits or qubit pairs~\cite{}.

\begin{figure}[t]
    \centering
    \includegraphics[width=1.0\columnwidth]{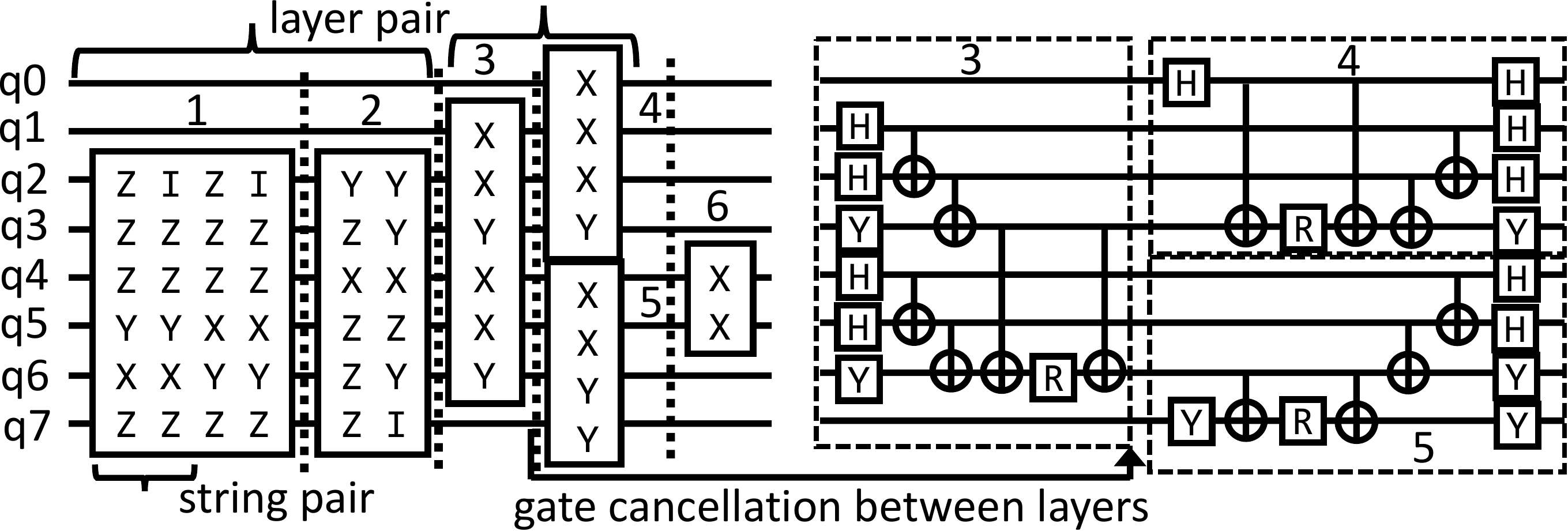}
    \vspace{-15pt}
    \caption{Example of compilation onto FT backend}
    %\AJ{gap on wires between left and right circuits}}    
    %\ANBANG{change final XY to XX on qubit4 and 5, and eliminate final CNOT, H, H on right hand}
    %\vspace{-5pt}
    \label{fig:FTexample}
\end{figure}

%The objective of our compiler pass for FT backend 
Our strategy for the FT backend %is to
is to adaptively find the synthesis plan 
%for all Pauli strings 
that can maximize gate cancellation %among them
since the mapping overhead can usually be neglected after applying quantum error correction~\cite{fowler2012surface}.
%We propose a circuit synthesis and gate cancellation algorithm for the fault tolerant backend.
The pseudocode % of %our proposed 
%optimization algorithm 
is shown in Algorithm~\ref{alg:FTsynthesis}, and we explain it with 
%the example in
Figure~\ref{fig:FTexample}.
To capture the  major gate cancellation opportunities, we scan over all layered blocks and try to select consecutive layer pairs that share the most Pauli operators. 
%These $pauli\_block$s have been rescheduled in a layered structure and 
There should be significant operator overlap between consecutive layers since this was considered in our scheduling.
The blocks on the left of Figure~\ref{fig:FTexample} are in five layers. 
Layer 1, 2, 3, 5 have one block in each and layer 4 has two blocks.
%The first three layers and the last layer have one block in each of them and the fourth layer has two blocks.
We will pair the layer 3 and 4 % third and the fourth layers 
together first since they share the same Pauli operators on 6 qubits.
Then the first two layers are paired since they share Pauli operators on only 2 qubits. %\AJ{they share 6 qubits?}.
The last layer is left alone.

We first realize gate cancellation between the paired layers.
%For all the paired layers, we %first
For all layer pairs, we
synthesize the Pauli strings at the end of the first layer and the  Pauli strings at the beginning the second layer in the pair. %since we already know that they share common Pauli operators.
For the layer 3 (block 3) and the layer 4 (block 4 and 5), we need to handle the $IYXXYXXI$ in the layer 3 and ($IIIIYXXX$,$YYXXIIII$) in the layer 4.
There are two sets of overlapped operators, $YXX$ on qubit 3-1 and $YXX$ on qubit 6-4.
For each set, most gates can be directly cancelled, and we can select one qubit from each set and connect them with CNOT gates.
The synthesis result for these two layers with gates cancelled is shown on the right of Figure~\ref{fig:FTexample}.
We repeat this process to optimize the synthesis of Pauli strings at the junction of two paired layers.
Here we will synthesize the last string in layer 1 and the first string in layer 2.
\begin{algorithm}[t]
\SetAlgoLined
 %\SetKwFor{For}{for}{do}{}
\KwIn{List of Pauli layers $pls$}
\KwOut{A quantum circuit of basic gates}

 %$pls = $list of scheduled Pauli layers\; 
 $pl\_paired=[]$; \tcp{paired Pauli layer list}
 
 \While{neighboring layers exist in $pls$}{
  $i = \mathrm{argmax}_{i\in IndexSet(pls)} Overlap(pl_i, pl_{i+1})$ \;
 $pls.remove(pl_i); \ pls.remove(pl_{i+1})$\;
 $pl\_paired.append((pl_{i}, pl_{i+1}))$\;
 }

\For{$(pl_1, pl_2)$ in $pl\_paired$}{
    $ps\_list_1 = $ last Pauli string of $pl_1$\;
    $ps\_list_2 = $ first Pauli string of $pl_2$\;
    %pb_2[0]$\; \tcp{First Pauli string of $pb_2$}
    analyze string overlap then do synthesis on $(ps\_list_1,ps\_list_2)$\;
    $pl_1.remove(ps\_list_1)$; $pl_2.remove(ps\_list_2)$\;
    \For{$pb$ in $(pl_1 + pl_2)$}{
        most\_overlap\_sort($pb$); \tcp{find overlap at Pauli-string-level}
        analyze string overlap then do synthesis on the sorted strings in $pb$\;
    }
}
%\tcp{Handle remaining Pauli blocks in $pls$}
\For{$pb$ in $pls$}{
    most\_overlap\_sort($pb$); analyze string overlap then do synthesis on the sorted strings in $pb$\;
}

%\v%space{-10pt}
\caption{Optimization for FT backend}
\label{alg:FTsynthesis}
\end{algorithm}

We then realize the gate cancellation between strings inside a block.
For those Pauli strings in the paired layer but not synthesized (one block with multiple strings), we employ a similar strategy at the string level for all Pauli strings inside one block.
For each block, we search for string pairs that share the same Pauli operators on the most qubits and then synthesize these pairs first.
%For example, i
In the block 1 in Figure~\ref{fig:FTexample}, the first three Pauli strings are not yet synthesized. 
We will pair and synthesize the first two Pauli strings since they share 5 Pauli operators and a lot of gates can be cancelled.
For the individual Pauli strings left, they are not paired with other strings (e.g., the third string in block 1).  
We check if it shares more Pauli operators with its left neighbor string or right neighbor string.
Then we select the one with more gate cancellation and synthesize the Pauli string accordingly.
For the blocks that are not paired with other blocks at the beginning of this algorithm (e.g., block 6% in Figure~\ref{fig:FTexample}
), we treat them as unsynthesized Pauli strings and apply the same strategy, pairing and synthesizing the strings with high gate cancellation potential first then dealing with individual strings.
Finally, all Pauli strings are compiled and we obtain a gate sequence of the input Pauli IR program.
The final gate count is substantially reduced because the gate cancellation potential created by our block scheduling passes is maximally exploited through the adaptive synthesis plan in our block-wise optimization pass.

%\AJ{Re-emphasize that the lexicographical ordering discussed in Section 4 created the potential for this adaptive synthesis plan to be maximally effective.}

% \AJ{Near-Term}
\subsection{On the Near-Term Superconducting Backend}

The compilation is more complicated for the SC backend because %the CNOT gates are supported only on neighboring physical qubits and 
the SWAP gates are necessary to change the qubit mapping due to the qubit connectivity constraints. 
The gates are not uniform as they have different error rates on different qubits.
We assume that the device calibration information (qubit coupling graph and the gate error rates on each qubit and qubit pair) is 
provided by the vendor.
The major objective  on the SC backend is to reduce the mapping overhead.
%number of SWAPs inserted during qubit routing.
%We need to determine the qubit initial layout. 
%Then we construct the CNOT tree and select the root qubit for each Pauli string in the Pauli IR program.

\begin{figure}[t]
    \centering
    \includegraphics[width=1.0\columnwidth]{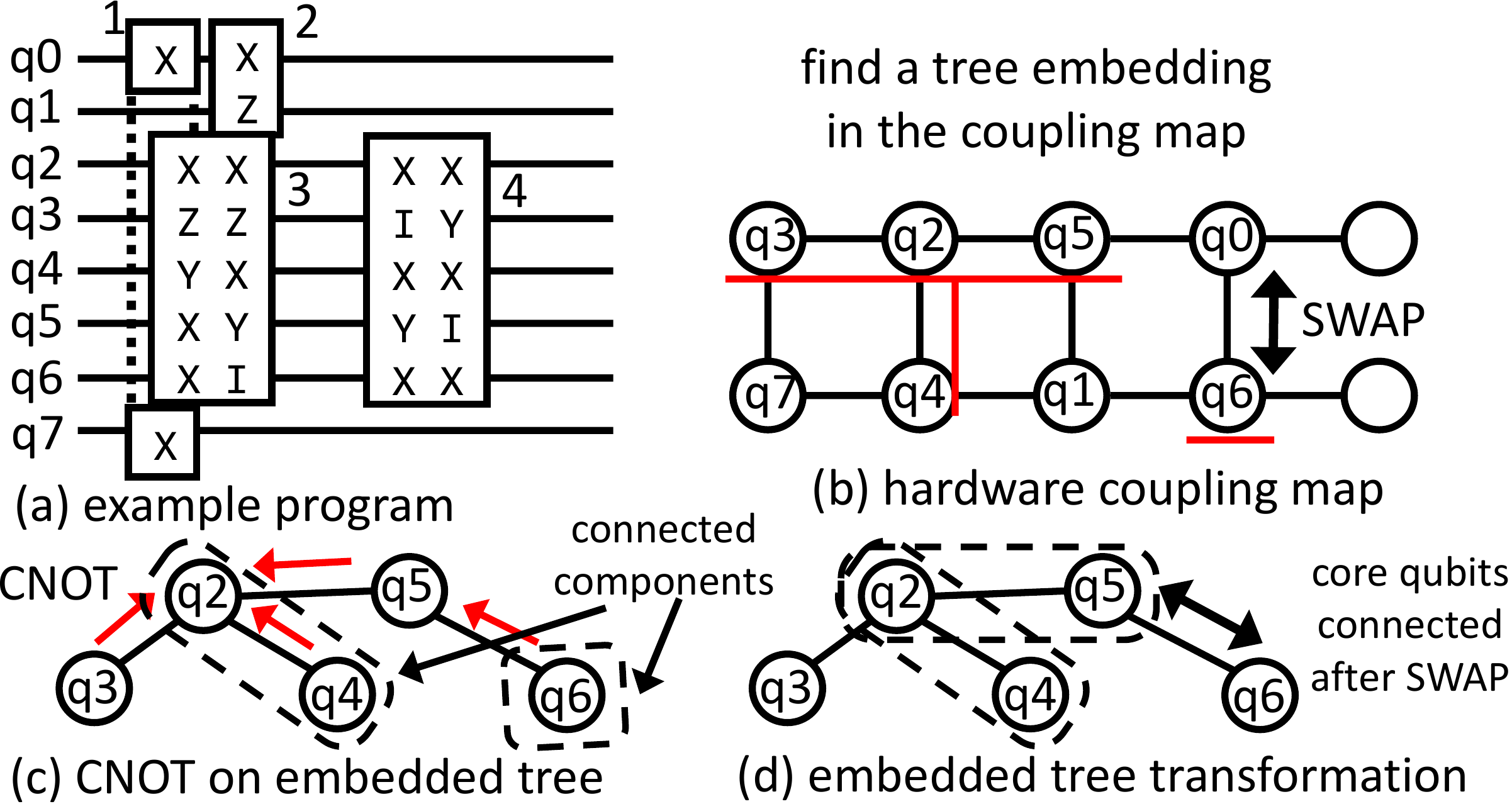}
    \vspace{-15pt}
    \caption{Example of compilation onto SC backend}
    \vspace{-5pt}
    \label{fig:SCsynthesis}
\end{figure}

%The key idea in this pass
Our key idea
is to find a tree embedding in the coupling map that can support the Pauli strings in the current layer and also minimize the mapping transition overhead between layers. 
Algorithm~\ref{alg:SCsynthesis} shows the pseudocode, and we explain it using the example in Figure~\ref{fig:SCsynthesis}.
For the initial qubit layout, we map all qubits to the most connected subgraph in the device coupling map.
%We do not consider specific gates since the synthesis of Pauli string circuits only requires that all qubits are in a connected subgraph and we do not have any gates in Pauli IR.
Suppose the coupling map and the current mapping of Figure~\ref{fig:SCsynthesis} (b).
We then begin to generate the simulation circuits and insert SWAPs for the blocks that appear in the critical path.
In our block scheduling, we have already placed the blocks in different layers.
In each layer, the largest block (involving the most qubits) is most likely on the critical path.
Our optimization pass will first process the largest block in each layer, %and then handle 
followed by
the small blocks remaining.
The program in Figure~\ref{fig:SCsynthesis} (a) has two layers in which block 3 and 4 are the largest blocks.
%Block 3 and 4 are the largest blocks in the two layers.

%\setlength{\textfloatsep}{\textfloatsepsave}
%\setlength{\floatsep}{\floatsepsave}

%\todo{commute figure 9}
For each block, we first select a root qubit.
We define that the core qubit list of a block contains the qubits which have a non-identity operator on all Pauli strings in the block (e.g., $q$2-5 for block 3, $q$(2,4,6) for block 4).
%We then order the qubits by the number of Pauli strings in which the qubit has non-identity operators in the current block and the previous block.
For block 3, since it is the first layer, we  only need to consider itself.
For $q$2-5 in its core list, they are already in a connected subgraph (Figure~\ref{fig:SCsynthesis} (b)).
We select any one of them (e.g., $q$2) as the root.
And we only need to attach $q$6 to this subgraph by connecting it to any node of this graph.
Suppose we swap $q$6 with $q$0 and now all active qubits in this block  are connected in a
subgraph.
Active qubits are those qubits that have a non-identity operator in at least one string in this block.
We can naturally generate an embedded tree from the coupling map (Figure~\ref{fig:SCsynthesis} (c)).
%Since all Pauli strings simulation circuits 
%that cann
%embedded tree (naturally generated from the coupling map) as shown in %Figure~\ref{fig:SCsynthesis} (c).

\begin{algorithm}[t]
\SetAlgoLined
 
\KwIn{List of Pauli layers $pls$, device information}
\KwOut{Hardware compatible circuit Q}
Map logical qubits to the most connected subgraph of the device coupling map; \tcp{Initial mapping}

%$pls = $ All scheduled pauli layers\; 

%$s_{in} = \varnothing$\; %\tcp{interior set of current pauli block}
%$s^l_{in} = \varnothing$\; \tcp{interior set of previous pauli layer}
%$s_{out} = \varnothing$\; %\tcp{outer set of current pauli block}
%$s^l_{out} = \varnothing$\; \tcp{outer set of previous pauli layer}
%$s^l_{in} = \varnothing$, $s^l_{out} = \varnothing$; \tcp{interior/outer set of previous pauli layer}
%$remain\_layers = \varnothing$\; %\tcp{Layers of blocks after first synthesis pass}

\For{$pauli\_layer$ in $pls$}{
   $pb = $ the largest Pauli block in $pauli\_layer$\;
%   %$s_{out} = compute\_outer\_set(pb)$\;
%   $s_{in} = (s_{in}\cap s^l_{in})\cup (s_{in}\cap s^l_{out})\cup (s_{in}- s^l_{out})$\; \tcp{sort nodes in $s_{in}$; $s^l_{in}$ and $s^l_{out}$ is from the previous layer }
%   $s_{out} = s_{in} \cup ((s_{out}-s_{in})\cap s^l_{in})\cup ((s_{out}-s_{in})\cap s^l_{out})\cup ((s_{out}-s_{in})- s^l_{out})$\; %\tcp{sort nodes in $s_{out}$}
%    $s^l_{in} = s_{in}$,\, $s^l_{out} = s_{out}$\;
    $s = $ core qubit list of $pb$\;
    %\tcp{decide root node}
    $T_1 =$ node in $s$ %$s_{out}$ 
    with largest connected component\;
    connect active qubits in $pb$
    % $s_{out}$ 
    to tree $T_1$ through shortest path (lowest error rate)\;
    $wl = $leaves of $T_1$ sorted by depth\;
    \For{ps in pb}{
        \While{$wl \ne \varnothing$}{
            $n = wl.deque()$;\, $np = n.parent$\;
            \lIf {$n$ is the root of $T_1$}{continue}
            \uIf{$ps[n]\ne I$ and $ps[np] \ne I$}{
                add single-qubit gates based on $ps[n]$ and $ps[np]$; $Q.append(CNOT(n, np))$\;
            }
            \ElseIf{$ps[n]\ne I$ and $ps[np] == I$}{
                $Q.append(SWAP(n, np))$\;
            }
            $wl.append(np)$\;
        }
        generate the right half circuit of $ps$ reversely\;
    }
    
    \For{$spb$ in remaining blocks of $pauli\_layer$}{
%        $s_{in} = compute\_interior\_set(spb)$\;
%        $s_{out} = s_{in}\cup (compute\_outer\_set(spb)-s_{in})$\;
        $T_2$ = try\_construct\_tree($spb$); \tcp{Return NULL if changes $T_1$}
        synthesize $spb$ with $T_2$ if $T_1$ not changed; otherwise add $spb$ to $remain\_layers$\;
        %\uIf{$T_2$==NULL}
        %{
        %    $remain\_layers.append([spb])$\;}
        %\Else {
        %    synthesize $spb$ with $T_2$\;    
        %}
    } %\tcp{Handle remaining small Pauli blocks}
}
\While{$remain\_layers$ is not empty}{
Sort $remain\_layers$ by cumulative distance between active qubits\;
Synthesize first layer of $remain\_layers$ with the same strategy and remove it from $remain\_layers$\;
}
%\v%space{-10pt}
\caption{Optimization for SC backend}
\label{alg:SCsynthesis}
\end{algorithm}

Next we can synthesize the %Pauli 
strings in block 3.
The key idea is to naturally implement the CNOT tree in the Pauli circuits on the embedded tree so that we do not need to insert SWAPs for all individual CNOTs. %from the hardware coupling.
%After we put all the active qubits in the current block together, we now synthesize the CNOT tree and insert SWAPs.
%A embedded tree in the coupling map can naturally be generated from the active qubits in the current block since they have already been in a connected component.
%After swap $q6$ with $q0$ in Figure~\ref{fig:SCsynthesis} (b), the embedded tree is in (c).
%Now we generate the CNOT tree and the single-qubit gates from the outermost qubits to the root for all the Pauli strings in the current block.
%Our algorithm will
We generate CNOT gates and single-qubit gates from the outermost qubits to the root for all the Pauli strings in the current block.
If a qubit is active in the current Pauli string, we will check if its parent node is also active in the current Pauli string.
If so, we insert a CNOT between the qubit and its parent. 
Otherwise, we swap it with its parent so that the qubit can get closer to the root and will be connected by CNOT later.
In Figure~\ref{fig:SCsynthesis} (c), the generated $\rm CNOT$s are labeled by red arrows.
After we determine the left CNOT tree, the right CNOT tree can be generated by reversing the order of $\rm CNOT$s in the left tree.

%Such algorithm design can greatly reduce the number of $\rm SWAP$s since we adaptively select an embedded tree in the hardware coupling map that fits the Pauli strings in the current block.
%A lot of gates can also be naturally canceled  since consecutive Pauli strings (which have been scheduled in a lexicographic order in Section~\ref{sec:interblock}) share many Pauli operators.

After we process block 3, we will %need to
compile block 4, the next block in the critical path.
%When selecting the root qubit from this list, we hope to minimize the transition overhead by selecting the qubits that are both active in the current block (block 4) and the previous block (block3).
As our block scheduling passes tend to maximize the overlap between two consecutive layers, the core lists of two consecutive layers are similar.
For example, $q$(2,4,6) are in the core list of block 4 and they all appear in the core list of  block 3.
We evaluate all these qubits to select the root qubit with the largest connected component (within the core list) in the current mapping (Figure~\ref{fig:SCsynthesis} (c)) to minimize the transition overhead.
For $q$(2,4,6), 
%if the mapping is like Figure~\ref{fig:SCsynthesis} (b), 
we will select $q2$ or $q4$ since they are in a size-2 connected component while $q6$ is in a size-1 connected component.
%we will select $q2$ or $q4$ since they are in a size 2 connected component while $q6$ only has size 1 connected component.
%Figure~\ref{}
Similarly, we then move all other active qubits to the tree %, which have non-identity operator on at least one Pauli string in the block, to the tree %with root
through the path with the smallest error rate estimated by the device information.
%estimated by the device information) on the coupling map.
%Here the shortest path indicates the path with smallest error rate estimated by the device error information.
Here we select $q2$ as the root and then swap $q6$ and $q5$ to transit from block 3 to 4 with only 1 SWAP (Figure~\ref{fig:SCsynthesis} (d)).
After that the core qubits in block 4  are connected and we can begin synthesizing 
%the circuit for 
all %Pauli 
strings in block 4.

The procedure above is to process the largest blocks in each layer.
For other small blocks in the same layer, we follow a similar strategy and attempt to construct the trees for active qubits in those small blocks.
If the trees of the small blocks do not affect the tree construction of the large block, we just process the small blocks in parallel with the large block since they will not affect each other.
This will create parallelism and  reduce the depth of the generated circuit. 
For example, block 2 and 3 can be processed in parallel because $q0$ and $q1$ are connected after swapping $q6$ with $q0$.
However, if the trees of the small blocks affect the processing of the large block, we will put it in $remain\_layer$ and process them at the end.
For example, block 1 will be in the $remain\_layer$ because connecting $q0$ and $q7$ will affect block 3.

After we process all the large blocks in the critical path and those small blocks that can executed in parallel, we will compile the blocks in the $remain\_layer$.
The order of processing these blocks is determined by whether the active qubits are close in the current mapping.
We compute the cumulative distance between active qubits in a block and then compile the block with the smallest cumulative distance and update the qubit mapping.
This process is repeated until all the blocks are processed and the compilation finishes.

\section{Evaluation}\label{sec:evaluation}

In this section, we evaluate \myCompilerNameSpace by comparing with state-of-the-art baselines, analyze the effects of individual passes, and perform real system study.
%we perform a comprehensive evaluation for the optimizations in \myCompilerNameSpace on various \QSK s and backends.

\subsection{Experiment Setup}

%\textbf{Baseline:}

\textbf{Backend:} 
The optimizations in this paper target two different backends, the fault-tolerant backend (FT) and near-term superconducting backend (SC). 
We will cover both of them. % in our evaluation.
We select IBM's latest 65-qubit Manhattan architecture~\cite{chamberland2020topological} as the SC backend. % architecture.
For real system study, we use IBM's 16-qubit Melbourne chip, the largest publicly available one.

\textbf{Metric:}
We use the CNOT/single-qubit gate count, and the circuit depth in the post-compilation program to evaluate \myCompilerName.
For the SC backend, the CNOT gate count is more important due to its higher error rate and latency. The depth is also important due to short qubit coherence times.
For the FT backend, T gate is usually more expensive but for the simulation kernels, the ratio between the H, Y, CNOT gate count and the T gate count grows linearly as the number of qubits increases.
Because a Pauli string of length $n$ will have $O(n)$ H, Y, and CNOT gates but the number of Rz gates (the only source of T gates) is always one.
It has also been shown that CNOT count is a significant cost in fault-tolerant algorithms and should not be neglected compared to T gates~\cite{maslov2016optimal}.
%but exact T gate counts can only be determined after compilation to the Clifford+T gate set (a common gate set for FT backends), but this in turn relies on the time step length $\Delta t$ which is not determined. 
Hence, we estimate the performance with total gate count and circuit depth, following convention in previous work~\cite{van2020circuit,hastings2015improving,cowtan2019phase,cowtan2020generic}. 
%we care more about the circuit depth since a FT is expected to have rich hardware resources to execute many gates in parallel and the circuit depth can indicate the execution time improvement 
%\AJ{since you already have a comprehensive evaluation in Tables 2 & 3, do you need to say anything here about which is a more important metric? It gives reviewers something to complain about since e.g. most current IBM gates are coherence limited and so optimizing depth becomes important. Conversely FT overheads can easily be dominated by T count}.
%A more detailed study requires the implementation 
%and 
%modeling 
%of a quantum error correction protocol, which is left as future work intentionally.
%For experiments on real systems, we use the probability of measuring the correct output to evaluate the optimization improvement.

%and the estimated successful probability, which is a theoretical estimation of the successful probability based on the input program and the hardware noise model and has been widely used in guiding the compiler optimization~\cite{nishio2020extracting,tannu2019ensemble,murali2019noise}. 
%We also include the probabilities of measuring the correct output.

%For SC, we have experiments for QAOA-MC on \todo{} and we use the probability of measuring the correct output to evaluate the overall improvement of \myCompilerName.

\textbf{Benchmark:} %\todo{}
We select 31 benchmarks of different sizes and various applications. 
For the  SC backend, we select VQE UCCSD ansatzes~\cite{peruzzo2014variational} of six sizes, and the QAOA programs~\cite{farhi2014quantum} for graph max-cut on regular (REG)  graphs of degrees 4, 8, 12,  and random (Rand) graphs of edge probability 0.1, 0.3, 0.5, as well as traveling salesman problem (TSP) of different sizes. 
These benchmarks are generated by Qiskit~\cite{Qiskit}.
For the FT backend, we first generate the Hamiltonians of five %chemical 
molecules using PySCF~\cite{PYSCF} ($\rm N_2$, $\rm H_2S$, $\rm MgO$, $\rm CO_2$, $\rm NaCl$).
We also prepare the Hamiltonians of three Ising models and three Heisenberg models, both of which are widely used in condensed matter physics, of different dimensions.
We finally generate random Hamiltonians (Rand) of various sizes (30 to 80 qubits) for a more comprehensive evaluation.
For a Hamiltonian of $n$ qubits, we prepare $5n^2$ \ps s. % of length $n$. 
In each \ps, we first randomly select one integer $m$ between $1$ and $n$. Then we randomly select $m$ qubits and assign random Pauli operators to them. 
The rest $n-m$ qubits will be assigned with the identity. 
Table~\ref{tab:benchmark} shows the details of these benchmarks.
Note that `Pauli \#' represents the number of \ps s. % in the corresponding Pauli IR program.
%, including the number of qubits and the number of \ps s when expressing the corresponding \QSK s in the Pauli IR.
We include the $\rm CNOT$ and single-qubit gate counts when naively converting these benchmarks into gates without any optimization/transformations,
and neglecting mapping overhead.

\begin{table}[t]
\center\vspace{-5pt}
\caption{Benchmark information}
\vspace{-5pt}
\resizebox{0.75\columnwidth}{!}{\begin{tabular}{|c|c|c|c|c|c|c|}
\hline
Backend & Type                      & Name      & Qubit \# & Pauli \# & CNOT \# & Single \# \\ \hline
\multirow{14}{*}{SC} & \multirow{6}{*}{UCCSD}    & UCCSD-8       & 8       & 144       & 1134    & 1240      \\ \cline{3-7}  
&                          & UCCSD-12      & 12       & 1476      & 16192   & 15588     \\ \cline{3-7} 
&                          & UCCSD-16        & 16       & 4200       & 56558     & 47044       \\ \cline{3-7} 
&                          & UCCSD-20      & 20       & 8316     & 132326  & 109248 \\ \cline{3-7} 
&                          & UCCSD-24        & 24       & 9300       & 146312     & 115584       \\ \cline{3-7} 
&                          & UCCSD-28      & 28       & 20724     & 353984  & 270196    \\ \cline{2-7}
&\multirow{8}{*}{QAOA}     & REG-20-4    & 20       & 40       & 80      & 40        \\ \cline{3-7} 
&                          & REG-20-8        & 20       & 80       & 160      & 80        \\ \cline{3-7} 
&                          & REG-20-12        & 20       & 120       & 240      & 120        \\ \cline{3-7} 
&                          & Rand-20-0.1        & 20       & 18       & 37      & 18        \\ \cline{3-7}  
&                          & Rand-20-0.3        & 20       & 56       & 113      & 56        \\\cline{3-7} 
&                          & Rand-20-0.5        & 20       & 93       & 187      & 93        \\ \cline{3-7} 
&                          & TSP-4        & 16       & 112       & 192      & 112        \\ \cline{3-7}  
&                          & TSP-5       & 25      & 225     & 400    & 225      \\ \hline  %\hline
\multirow{14}{*}{FT} & \multirow{3}{*}{Ising}    & Ising-1D  & 30       & 29       & 58      & 29        \\ \cline{3-7}  
&                          & Ising-2D & 30       & 49        & 98      & 29        \\ \cline{3-7} 
&                          & Ising-3D  & 30       & 59       & 118      & 59       \\ \cline{2-7}
&\multirow{3}{*}{Heisenberg}    & Heisen-1D  & 30       & 87       & 174      & 319        \\ \cline{3-7} 
&                          & Heisen-2D & 30       & 147        & 294      & 539        \\ \cline{3-7} 
&                          & Heisen-3D  & 30       & 177       & 354      & 649       \\ \cline{2-7}
&\multirow{5}{*}{Molecule} 
                          & $\rm N_2$ & 20       & 2951    & 39594 & 32151   \\ \cline{3-7} 
&                          & $\rm H_2S$       & 22       & 4582     & 66026   & 52686     \\ \cline{3-7} 
&                          & $\rm MgO$       & 28       & 24239     & 388258   & 310519     \\ \cline{3-7} 
&                          & $\rm CO_2$       & 30       & 16154     & 252402   & 202282     \\ \cline{3-7} 
&                          & $\rm NaCl$       & 36       & 67667     & 1249768   & 945935     \\ \cline{2-7}
&\multirow{6}{*}{Random} 
& Rand-30      & 30       & 4500     & 132939   & 99123     \\ \cline{3-7} 
& & Rand-40       & 40       & 8000     & 316039   & 229240     \\ \cline{3-7} 
& & Rand-50       & 50       & 12500     & 618763   & 441532     \\ \cline{3-7} 
& & Rand-60       & 60       & 18000     & 1068153   & 754071     \\ \cline{3-7} 
& & Rand-70       & 70       & 24500     & 1699771   & 1190101     \\ \cline{3-7} 
& & Rand-80       & 80       & 32000     & 2540640   & 1768117     \\
\hline
\end{tabular}}
\vspace{-5pt}
\label{tab:benchmark}
\end{table}

\textbf{Implementation:} We prototype \myCompilerNameSpace in Python 3.8 (denoted by `PH').
The entire compilation flow has two stages.
The first stage is the quantum simulation program optimizations.
The baselines include the CQC $\rm t|ket\rangle$ compiler~\cite{sivarajah2020t} which employs the simultaneous diaganolization~\cite{ cowtan2019phase, cowtan2020generic, de2020architecture}, a popular technique for optimizing quantum simulation programs (`TK'), and the QAOA compiler~\cite{alam2020circuit,alam2020efficient,alam2020noise}, an algorithm-specific compiler for unconstrained optimization QAOA on graphs (`QAOA compiler').
The second stage is the generic compilation and we have two industry generic compilers, the IBM's Qiskit~\cite{Qiskit} at the highest optimization level 3 (`Qiskit\_L3') and the CQC  $\rm t|ket\rangle$ generic compiler~\cite{sivarajah2020t} at the highest optimization level 2 (`tket\_O2'), 
The experiments are performed on a server with a 28-core Intel Xeon Platinum 8280 CPU and 1TB RAM.
Note that due to the limited representation ability of $\rm t|ket\rangle$, the algorithmic constraints are hard to be encoded in `TK'. 
To run our experiments and perform a fair comparison at our best, we relax those constraints in `TK' and this relaxation  allows a larger optimization space.

\subsection{Comparing with $\rm t|ket\rangle$ and the QAOA Compiler}

\begin{table*}[t]
  \centering
  \caption{Compilation time and results compared with $\rm t|ket\rangle$~\cite{sivarajah2020t}}
  \vspace{-5pt}
  \resizebox{\textwidth}{!}{
    \begin{tabular}{|c|c|c|c|c|c|c|c|c|c|c|c|c|c|c|c|c|c|c|c|c|c|c|}
    \hline
        &     \multicolumn{2}{c|}{Time(s)}  & \multicolumn{4}{c|}{PH+Qiskit\_L3}    & Time(s) & \multicolumn{4}{c|}{PH+tket\_O2}      &   \multicolumn{2}{c|}{Time(s)}   & \multicolumn{4}{c|}{TK+Qiksit\_L3}  & Time(s) & \multicolumn{4}{c|}{TK+tket\_O2} \\
     \hline
          & PH  & Qiskit  & CNOT  & Single & Total & Depth & tket & CNOT  & Single & Total & Depth & TK  & Qiskit & CNOT  & Single & Total & Depth & tket & CNOT  & Single & Total & Depth \\

    %      &       & \multicolumn{5}{c|}{PH+Qiskit\_L3}    & \multicolumn{5}{c|}{PH+tket\_O2}      &       & \multicolumn{5}{c|}{TK+Qiksit\_L3}  & \multicolumn{5}{c|}{TK+tket\_O2} \\
  % \hline
   %       & PH time & Qiskit time  & CNOT  & Single & Total & Depth & tket\_O2 time & CNOT  & Single & Total & Depth & TK time & Qiskit time & CNOT  & Single & Total & Depth & tket\_O2 time & CNOT  & Single & Total & Depth \\
   \hline
    UCCSD-8 & 0.5   & 7     & 1160  & 669   & 1829  & 1382  & 16    & 1228  & 535   & 1763  & 1367  & 0.5   & 13    & 2187  & 1180  & 3367  & 1928  & 1     & 1723  & 346   & 2069  & 1499 \\
    \hline
    UCCSD-12 & 5     & 98    & 17498 & 8633  & 26131 & 17741 & 485   & 20023 & 6796  & 26819 & 18665 & 1     & 274   & 35775 & 16054 & 51829 & 28183 & 39    & 25927 & 4008  & 29935 & 21294 \\
    \hline
    UCCSD-16 & 17    & 379   & 57247 & 34235 & 91482 & 56226 & 7300  & 83244 & 18767 & 102011 & 69332 & 2     & 1297  & 142053 & 58569 & 200622 & 107617 & 245   & 104185 & 11838 & 116023 & 86744 \\
    \hline
    UCCSD-20 & 37    & 788   & 107372 & 68691 & 176063 & 100352 & 11033 & 132644 & 38179 & 170823 & 109685 & 5     & 2908  & 324258 & 129678 & 453936 & 240743 & 903   & 221051 & 23739 & 244790 & 179108 \\
   \hline
    UCCSD-24 & 43    & 982   & 124340 & 76946 & 201288 & 117701 & 33155 & 188851 & 39858 & 228709 & 152422 & 6     & 3283  & 387570 & 158810 & 546380 & 291619 & 1331  & 256610 & 29099 & 285709 & 207850 \\
   \hline
    UCCSD-28 & 103   & 2461  & 290829 & 179186 & 470015 & 260925 & 69940 & 487616 & 96019 & 583635 & 396007 & 15    & 8682  & 875029 & 339854 & 1214883 & 644263 & 3456  & 598686 & 58785 & 657471 & 486507 \\
   \hline
    REG-20-4 & 0     & 0.14  & 366   & 128   & 494   & 147   & 0.27  & 382   & 43    & 425   & 152   & 0     & 0.28  & 1513  & 494   & 2007  & 760   & 0.18  & 1378  & 40    & 1418  & 667 \\
    \hline
    REG-20-8 & 0     & 0.23  & 539   & 254   & 793   & 246   & 0.56  & 705   & 81    & 786   & 313   & 0     & 0.4   & 1858  & 647   & 2505  & 860   & 0.18  & 1721  & 80    & 1801  & 812 \\
    \hline
    REG-20-12 & 0     & 0.29  & 678   & 354   & 1032  & 319   & 1.08  & 959   & 120   & 1079  & 412   & 0     & 0.3   & 1858  & 660   & 2518  & 783   & 0.15  & 1678  & 120   & 1798  & 764 \\
    \hline
    Rand-20-0.1 & 0     & 0.08  & 104   & 67    & 171   & 56    & 0.13  & 188   & 20    & 208   & 81    & 0     & 0.2   & 520   & 183   & 703   & 241   & 0.06  & 434   & 19    & 453   & 205 \\
   \hline
    Rand-20-0.3 & 0     & 0.1   & 398   & 186   & 584   & 192   & 0.46  & 520   & 58    & 578   & 228   & 0     & 0.24  & 1504  & 496   & 2000  & 691   & 0.18  & 1324  & 57    & 1381  & 626 \\
   \hline
    Rand-20-0.5 & 0     & 0.12  & 550   & 302   & 852   & 266   & 1.14  & 801   & 95    & 896   & 332   & 0     & 0.28  & 1712  & 617   & 2329  & 726   & 0.2   & 1605  & 94    & 1699  & 737 \\
    \hline
    TSP-4 & 0     & 0.25  & 434   & 245   & 679   & 239   & 0.81  & 712   & 112   & 824   & 333   & 0     & 0.21  & 1038  & 480   & 1518  & 479   & 0.09  & 2327  & 112   & 2439  & 505 \\
    \hline
    TSP-5 & 0     & 4.44  & 1179  & 604   & 1783  & 504   & 2.21  & 1573  & 225   & 1798  & 626   & 0     & 0.46  & 3022  & 1194  & 4216  & 1057  & 0.45  & 2467  & 225   & 2692  & 1052 \\
    \hline
    N2    & 10    & 205   & 15981 & 11366 & 27347 & 16788 & 71    & 16192 & 11069 & 27261 & 17524 & 3     & 94    & 19708 & 11294 & 31002 & 21196 & 13    & 18928 & 9702  & 28630 & 20446 \\
    \hline
    H2S   & 23    & 359   & 24792 & 17307 & 42099 & 26581 & 129   & 25243 & 16936 & 42179 & 28040 & 4     & 204   & 35210 & 20597 & 55807 & 36642 & 38    & 33968 & 17077 & 51045 & 35329 \\
    \hline
    MgO   & 523   & 1960  & 96831 & 67989 & 164820 & 93694 & 4917  & 116035 & 80175 & 196210 & 130558 & 18    & 1664  & 198158 & 115645 & 313803 & 205508 & 3903  & 192499 & 85693 & 278192 & 194908 \\
    \hline
    CO2   & 260   & 1387  & 98346 & 79122 & 177468 & 96588 & 6339  & 94927 & 61670 & 156597 & 96816 & 15    & 1037  & 126114 & 70731 & 196845 & 130690 & 1664  & 121768 & 56283 & 178051 & 124558 \\
   \hline
    NaCl  & 5621  & 6606  & 316472 & 258949 & 575421 & 338965 & 18731 & 307038 & 223261 & 530299 & 342775 & 66    & 6581  & 623671 & 339088 & 962759 & 630398 & 77534 & 605202 & 247540 & 852742 & 599320 \\
   \hline
    Ising-1D & 0     & 0.15  & 58    & 29    & 87    & 6     & 0.01  & 58    & 29    & 87    & 6     & 0.02  & 1.29  & 508   & 59    & 567   & 451   & 0.05  & 508   & 29    & 537   & 450 \\
   \hline
    Ising-2D & 0     & 0.21  & 98    & 49    & 147   & 18    & 0.02  & 98    & 49    & 147   & 18    & 0.02  & 0.91  & 306   & 79    & 385   & 220   & 0.04  & 306   & 49    & 355   & 219 \\
    \hline
    Ising-3D & 0     & 0.28  & 118   & 59    & 177   & 18    & 0.02  & 118   & 59    & 177   & 18    & 0.02  & 0.89  & 290   & 89    & 379   & 189   & 0.04  & 290   & 59    & 349   & 188 \\
   \hline
    Heisen-1D & 0.01  & 0.43  & 87    & 203   & 261   & 13    & 0.06  & 87    & 190   & 277   & 13    & 0.03  & 0.66  & 172   & 229   & 401   & 127   & 0.07  & 169   & 200   & 369   & 126 \\
   \hline
    Heisen-2D & 0.02  & 0.84  & 216   & 311   & 527   & 43    & 0.11  & 212   & 284   & 496   & 47    & 0.07  & 1.02  & 293   & 351   & 644   & 102   & 0.07  & 293   & 294   & 587   & 98 \\
    \hline
    Heisen-3D & 0.028 & 1.07  & 305   & 363   & 668   & 65    & 0.15  & 295   & 335   & 630   & 67    & 0.09  & 1.28  & 365   & 403   & 768   & 135   & 0.08  & 364   & 328   & 692   & 125 \\
   \hline
    Rand-30 & 25    & 1157  & 93885 & 56241 & 150126 & 70213 & 1173  & 88943 & 50523 & 139466 & 77704 & 6     & 544   & 112894 & 65291 & 178185 & 86930 & 130   & 107846 & 61720 & 169566 & 96682 \\
   \hline
    Rand-40 & 72    & 3455  & 238073 & 133152 & 371225 & 171664 & 42807 & 228725 & 121698 & 350423 & 194276 & 13    & 1613  & 276127 & 151661 & 427788 & 206974 & 1390  & 265359 & 144981 & 410340 & 235020 \\
   \hline
    Rand-50 & 151   & 7624  & 470924 & 253551 & 724475 & 327689 & >72hrs & \multicolumn{4}{c|}{N/A}     & 27    & 3121  & 532536 & 285004 & 817540 & 387974 & 10743 & 513447 & 275527 & 788974 & 443640 \\
   \hline
    Rand-60 & 309   & 14320 & 838322 & 438654 & 1276976 & 579384 & >72hrs & \multicolumn{4}{c|}{N/A}     & 48    & 6553  & 928918 & 487809 & 1416727 & 666906 & 25304 & 897772 & 473898 & 1371670 & 766627 \\
    \hline
    Rand-70 & 552   & 24257 & 1344276 & 690099 & 2034375 & 918406 & >72hrs & \multicolumn{4}{c|}{N/A}     & 80    & 8713  & 1475377 & 762111 & 2237488 & 1050286 & 111202 & 1428558 & 743649 & 2172207 & 1214547 \\
    \hline
    Rand-80 & 989   & 36869 & 2037292 & 1029742 & 3067034 & 1390099 & >72hrs & \multicolumn{4}{c|}{N/A}     & 126   & 18064 & 2207500 & 1130670 & 3338170 & 1562077 & >72hrs & \multicolumn{4}{c|}{N/A} \\
    \hline
    \end{tabular}%
  }\label{tab:tketcompare}%
\end{table*}%

Table~\ref{tab:tketcompare} shows the compilation time and results of the four configurations of all benchmarks on the two backends. 
Note that `>72 hrs' indicates that the `tket\_O2' takes over 72 hours and was shut down in the middle.
%The data for time and `PH+Qiskit\_3' are absolute while the data for the other three configurations are relative.
In summary, `PH' outperforms `TK' with substantial  gate count and circuit depth reduction while only introducing $\sim 5\%$ additional time (`PH' vs `Qiskit/tket time') in the entire compilation flow.
%The label 

On the SC backend, `PH' can reduce the CNOT, single-qubit, total gate count, and circuit depth by 66.2\% (43.3\%), 53.4\% (-22.7\%), 62.6\% (41.2\%), and 60.8\% (44.3\%), respectively on average, compared with `TK' using `Qiskit\_L3' (`tket\_O2') generic compilation.
%When using `tket\_O2', `PH' can reduce the CNOT, single-qubit, total gate count, and circuit depth by 43.3\%, -22.7\%, 41.2\%, and 44.3\%, respectively, compared with `TK'.
%`PH' can achieve such significant improvement because the mapping overhead is greatly reduced
PH' can achieve such significant improvement  because `TK' does not support mapping-aware optimization for general Pauli strings and can only do a inefficient generic qubit mapping.
%Such significant improvement comes from the BC pass in `PH' where the mapping overhead is significantly reduced.
%`TK' does not support mapping-aware optimization for general Pauli strings and the mapping is inefficiently done in the generic compilation. %, which is very inefficient.
The single-qubit gate count increases when using `tket\_O2' but this does not affect the overall improvement since the CNOT gates have much higher error rates on the SC backend and latency and the total single-qubit gate count is still relatively low.
%the single-qubit gates have relatively low error rates on the SC backend and the total single-qubit gate count is much smaller than the CNOT count.

On the FT backend, `PH' can reduce the CNOT, single-qubit, total gate count, and circuit depth by 38.7\% (44.5\%), 18.6\% (3.0\%), 32.8\% (34.4\%), and 61.7\% (68.0\%), respectively on average, compared with `TK' using `Qiskit\_L3' (`tket\_O2').
%When using `tket\_O2', `PH' can reduce the CNOT, single-qubit, total gate count, and circuit depth by 44.5\%, 3.0\%, 34.4\%, and 68.0\%, respectively on average, compared with `TK'.
The circuit depth reduction is significant due to the depth-oriented scheduling in `PH'. %\myCompilerName.
%We can see that the circuit depth reduction is significant due to the `DO' pass in \myCompilerName.
%Our BC pass is also very effective compared with the string cluster strategy in `TK' which will lose some gate cancellation opportunities.
Our block-wise optimization is also much effective compared with `TK' strategy.
The details of `TK' are not public and what we can infer, at our best, from their limited documents~\cite{sivarajah2020t, cowtan2019phase, cowtan2020generic, de2020architecture} is that the  simultaneous diagonalization may introduce too much overhead.
For example, the `Ising-1D' program has even more gates after `TK'. % compilation.
One possible reason is that all Pauli strings in Ising-1D are mutually commutative and it takes many additional gates to simultaneously diagonalize all these Pauli strings. 

%\textbf{Compilation time:}
%Table~\ref{tab:tketcompare} reports the execution time (in seconds) of the simulation kernel optimizations and the generic optimizations. 
%In general, the compilation time for Pauli string optimizations is much shorter than that of generic compilation.
%The average compilation time of `PH' is only 6.2\% and 5.1\% of that of `Qiskit\_L3' and `tket\_O2', respectively.
%$\rm t|ket\rangle$ at O2, respectively.
%The total compilation time is dominated by the generic compilation and `PH' only introduces a very small overhead.
%If we only consider the Pauli string optimizations, `PH' requires $4.5\times$ time compared with `TK'. % the Pauli string optimizations in $\rm t|ket\rangle$.
%This is because our \myCompilerNameSpace prototype is implemented in pure Python while  $\rm t|ket\rangle$ is highly optimized in C++. 
%We believe that this does not indicate any inefficiency of \myCompilerNameSpace due to the 
%since it is known that there exists a 
%huge intrinsic performance gap ($\sim 100\times$) between Python and C++.
%Finally, this difference is largely amortized by the very long generic compilation.

%\textbf{Comparing with the QAOA compiler~\cite{alam2020circuit}:}
%We also compare \myCompilerNameSpace with the QAOA compiler~\cite{alam2020circuit,alam2020efficient,alam2020noise} which on supports the QAOA on unconstrained optimizations.
Table~\ref{tab:QAOAcompiler} shows the compilation results of `PH' and the QAOA compiler~\cite{alam2020circuit} on the 6 MaxCut problems. 
We ran the QAOA compiler with 20 random seeds for each program and collected the averaged compilation results.
Comparing with the QAOA compiler, \myCompilerNameSpace can achieve 31.2\%, 16.3\%, and 24.1\% reduction in CNOT count, total gate count, and circuit depth, respectively on average, using only $1.7\%$ compilation time.
The overhead is about 50\% in single-qubit gate count, but in QAOA the CNOT count is usually over $3-4\times$ higher than single-qubit gate count and CNOT error rate is usually 10$\times$ higher on the SC backend.
Therefore, `PH' significantly outperforms QAOA compiler, even though it is more general purpose and not tailored to a single algorithm.
This is because `PH' employs a block-wise optimization for searching SWAPs and the search scope is much larger than that of the QAOA compiler's greedy search.
%while the QAOA compiler's search has relatively low depth and only considering few gates afterwards.
%Moreover, \myCompilerNameSpace is faster and the total compilation time (`PH+Qiskit\_L3') is averagely only 1.7\% of that of `QAOA compiler+Qiskit\_L3'.

%\subsection{Comparing with the QAOA compiler}
% Table generated by Excel2LaTeX from sheet 'Sheet2'
\begin{table}[t]
  \centering
   \vspace{-5pt}
  \caption{Comparing with QAOA compiler~\cite{alam2020circuit}}
  \vspace{-5pt}
      \resizebox{\columnwidth}{!}{ 
    \begin{tabular}{|c|c|c|c|c|c|c|c|c|c|c|}
    \hline
          & \multicolumn{5}{c|}{PH+Qiskit\_L3}    & \multicolumn{5}{c|}{QAOA\_Compiler+Qiskit\_L3} \\
    \hline
    Benchmark & CNOT  & Single & Total & Depth & Time(s)  & CNOT  & Single & Total & Depth & Time(s) \\
    \hline
    REG-20-4 & 366   & 128   & 494   & 147   & 0.14  & 394 & 101   & 495 & 171 & 6.32 \\
   \hline
    REG-20-8 & 539   & 254   & 793   & 246   & 0.23  & 727 & 141   & 868 & 297   & 10.27 \\
   \hline
    REG-20-12 & 678   & 354   & 1032  & 319   & 0.29  & 1020 & 181   & 1201 & 399 & 14.55 \\
   \hline
    Rand-20-0.1 & 104   & 67    & 171   & 56    & 0.08  & 212 & 80  & 292 & 111 & 4.52 \\
    \hline
    Rand-20-0.3 & 398   & 186   & 584   & 192   & 0.1   & 546 & 118 & 664 & 230 & 7.74 \\
   \hline
    Rand-20-0.5 & 550   & 302   & 852   & 266   & 0.12  & 842 & 155 & 997   & 334 & 12.3\\
   \hline
    \end{tabular}%
    }\vspace{-5pt}
  \label{tab:QAOAcompiler}%
\end{table}%

\subsection{Effect of the Passes}
Now we study the effect of the individual passes in \myCompilerName. We first compare the two block scheduling passes.
\iffalse
Table~\ref{tab:FTQCresults} shows the experiment results on the FT backend.
We apply all four configurations and their labels are in the first row. 
%`GCO' or `DO' represent the block scheduling option.
%`BC' indicates using the inner-block compilation in Section~\ref{sec:innerblock}.
We collect the CNOT gate count, single-qubit gate count, total gate count, and the circuit depth in the post-compilation gate sequence. 
We first introduce the overall improvement of different compilation options.
Then we discuss the effect of different Pauli string patterns in the benchmarks and how should choose the compilation optimization based on the program patterns in the \QSK s.
\fi

\textbf{DO vs GCO scheduling:} 
%We first compare the two block scheduling options. 
On the left of Table~\ref{tab:passeffect} we show the difference
%in compilation result 
between the depth-oriented (DO) scheduling and the gate-count-oriented (GCO) scheduling (in Section~\ref{sec:interblock}). %, introduced in Section~\ref{sec:interblock}.
%There is a trade-off between the circuit depth and the gate count.
Overall, across the 17 benchmarks on the FT backend, `DO' can yield low-depth circuits while `GCO' can reduce the gate count more. 
%the circuit depth of DO is only $48.1\%$ (geomean) compared with that of GCO while it requires $5.9\%$, $0.17\%$, and $3.2\%$ additional CNOT, single-qubit, and total gate count on average, respectively, when using the default circuit synthesis in Qiskit (`GCO+Qiskit\_L3' vs `DO+Qiskit\_L3').
%After applying BC, 
The circuit depth of DO is $46.7\%$ (geomean) compared with that of GCO and the gate count overhead is $5.9\%$, $0.64\%$, and $3.3\%$ for CNOT, single-qubit, and total gate count, respectively. % (`GCO+BC+ Qiskit\_L3' vs `DO+BC+Qiskit\_L3').
For benchmarks on the SC backend, the effect of the block scheduling is largely amortized by mapping overhead reduction since the tested Manhattan architecture has very sparse qubit connection.
For the UCCSD benchmarks, `DO' and `GCO' share similar overall performance.
For the QAOA benchmarks, there is no difference between `DO' and `GCO' since the entire kernel has only one block.
%DO achieves around $2\%\sim 3\%$ circuit depth reduction with $<1\%$ CNOT count variation. 

\begin{table}[t]
  \centering
  \caption{Effect of passes}
  \vspace{-5pt}
  \resizebox{1.0\columnwidth}{!}{
    \begin{tabular}{|c|c|c|c|c|c|c|c|c|}
    \hline
          & \multicolumn{4}{c|}{DO vs GCO} & \multicolumn{4}{c|}{Block-Wise Compilation improvement} \\
    \hline
          & CNOT  & Single & Total & Depth & CNOT  & Single & Total & Depth \\
    \hline
    UCCSD-8 & -3.89\% & -0.89\% & -2.82\% & 0.00\% & -28.97\% & -11.51\% & -23.44\% & -10.08\% \\
   \hline
    UCCSD-12 & 7.40\% & -8.13\% & 1.72\% & 1.98\% & -42.29\% & -30.92\% & -38.97\% & -32.32\% \\
    \hline
    UCCSD-16 & -4.19\% & -4.01\% & -4.12\% & -1.53\% & -52.90\% & -34.30\% & -47.32\% & -44.33\% \\
   \hline
    UCCSD-20 & -2.68\% & -8.59\% & -5.08\% & -4.22\% & -65.57\% & -44.17\% & -59.51\% & -59.94\% \\
    \hline
    UCCSD-24 & -5.88\% & -2.85\% & -4.74\% & -7.96\% & -63.08\% & -39.38\% & -56.59\% & -55.18\% \\
   \hline
    UCCSD-28 & 9.66\% & 5.27\% & 8.49\% & 4.38\% & -50.53\% & -19.09\% & -45.02\% & -49.80\% \\
   \hline
    REG-20-4 & N/A   & N/A   & N/A   & N/A   & -26.51\% & -23.81\% & -25.83\% & -11.45\% \\
    \hline
    REG-20-8 & N/A   & N/A   & N/A   & N/A   & -35.91\% & -21.85\% & -31.99\% & -19.34\% \\
   \hline
    REG-20-12 & N/A   & N/A   & N/A   & N/A   & -40.42\% & -25.94\% & -36.14\% & -22.57\% \\
   \hline
    Rand-20-0.1 & N/A   & N/A   & N/A   & N/A   & -55.93\% & -22.99\% & -47.06\% & -33.33\% \\
   \hline
    Rand-20-0.3 & N/A   & N/A   & N/A   & N/A   & -38.01\% & -21.19\% & -33.49\% & -16.88\% \\
    \hline
    Rand-20-0.5 & N/A   & N/A   & N/A   & N/A   & -42.29\% & -20.32\% & -36.04\% & -23.78\% \\
   \hline
    TSP-4 & N/A   & N/A   & N/A   & N/A   & -57.49\% & -40.10\% & -52.52\% & -23.15\% \\
   \hline
    TSP-5 & N/A   & N/A   & N/A   & N/A   & -54.64\% & -19.68\% & -46.79\% & -23.05\% \\
   \hline
    N2    & 10.60\% & 5.43\% & 8.39\% & -9.87\% & -8.28\% & -1.65\% & -5.63\% & -10.46\% \\
   \hline
    H2S   & 13.80\% & 4.78\% & 9.91\% & -6.54\% & -10.67\% & -4.67\% & -8.29\% & -12.65\% \\
   \hline
    MgO   & 25.85\% & 6.04\% & 16.84\% & -8.50\% & -4.97\% & -1.58\% & -3.60\% & -5.87\% \\
   \hline
    CO2   & 31.83\% & 10.45\% & 21.86\% & 2.42\% & -6.99\% & -7.79\% & -7.33\% & -9.72\% \\
   \hline
    NaCl  & 25.14\% & 6.46\% & 15.98\% & -5.25\% & -12.18\% & -9.79\% & -11.12\% & -13.29\% \\
    \hline
    Ising-1D & 0.00\% & 0.00\% & 0.00\% & -93.10\% & 0.00\% & 0.00\% & 0.00\% & 0.00\% \\
   \hline
    Ising-2D & 0.00\% & 0.00\% & 0.00\% & -68.42\% & 0.00\% & 0.00\% & 0.00\% & 0.00\% \\
    \hline
    Ising-3D & 0.00\% & 0.00\% & 0.00\% & -71.43\% & 0.00\% & 0.00\% & 0.00\% & 0.00\% \\
    \hline
    Heisen-1D & 0.00\% & -14.71\% & -10.31\% & -92.57\% & 0.00\% & 0.00\% & 0.00\% & 0.00\% \\
    \hline
    Heisen-2D & -19.10\% & -13.37\% & -15.81\% & -82.30\% & 0.00\% & -0.64\% & -0.38\% & 0.00\% \\
   \hline
    Heisen-3D & -8.41\% & -14.39\% & -11.76\% & -80.83\% & 0.00\% & 0.00\% & 0.00\% & 0.00\% \\
   \hline
    Rand-30 & 7.32\% & 6.55\% & 7.06\% & -9.92\% & -8.83\% & -4.75\% & -7.49\% & -29.71\% \\
   \hline
    Rand-40 & 5.74\% & 5.61\% & 5.70\% & -8.69\% & -9.72\% & -2.62\% & -7.57\% & -31.43\% \\
    \hline
    Rand-50 & 4.89\% & 4.79\% & 4.86\% & -9.14\% & -10.30\% & -3.18\% & -8.19\% & -32.79\% \\
    \hline
    Rand-60 & 4.05\% & 4.13\% & 4.07\% & -8.71\% & -10.74\% & -2.80\% & -8.45\% & -33.53\% \\
    \hline
    Rand-70 & 3.64\% & 3.53\% & 3.61\% & -8.78\% & -11.07\% & -1.70\% & -8.45\% & -33.97\% \\
   \hline
    Rand-80 & 3.29\% & 3.26\% & 3.28\% & -8.64\% & -11.12\% & -1.59\% & -8.49\% & -33.85\% \\
    \hline
    \end{tabular}%
    \vspace{-15pt}
  }\label{tab:passeffect}%
\end{table}%

\textbf{BC improvement:}
Our block-wise compilation (BC) passes (in Section~\ref{sec:innerblock}) can significantly reduce the gate count and circuit depth. % on both backends.
On the right of Table~\ref{tab:passeffect} we show the comparison between using BC against a naive synthesis and Qiskit\_L3.
%The BC optimizations can further reduce the gate count and circuit depth in both DO and GCO scheduling.
%When using the GCO scheduling, BC can reduce the circuit depth by $14\%$. The CNOT, single-qubit, and total gate counts are also reduced by $6.0\%$, $3.1\%$, and $5.0\%$, respectively (`GCO+Qiskit\_L3' vs `GCO+BC+Qiskit\_L3'). 
%For the DO scheduling, 
For the 17 benchmarks on the FT backend, BC reduces the circuit depth, the CNOT, single-qubit, and total gate counts  %are alreduced 
by $15.5\%$, $6.0\%$, $3.1\%$, and $5.0\%$, respectively.  %(`DO+Qiskit\_L3' vs `DO+BC+Qiskit\_L3'). 
On the SC backend, the BC pass is even more effective since the large mapping overhead can be greatly reduced.
For the UCCSD (QAOA) benchmarks, BC can %significantly
reduce the CNOT, single-qubit, total gate count, and circuit depth by  $54\%$ ($56\%$), $31\%$ ($20\%$), $47\%$ ($47\%$), and $45\%$ ($23\%$), respectively on average.
%For the QAOA benchmarks, BC can also reduce the CNOT, single-qubit, total gate count, and circuit depth by $56\%$, $20\%$, $47\%$, and $23\%$, respectively on average.
%\AJ{can you add a brief sentence about why QAOA compiler is not as effective? because that also exploits pauli strings.}
%The circuit depth is also reduced by round $23\%$ on average.

%The circuit depth is also reduced by round $45\%$ on average.

\textbf{Pauli string pattern effects:}
%We can observe that the 
It can be observed that the effect of the passes vary on different benchmarks.
The reason is that the Pauli strings in the benchmarks have different patterns which can be
%The Pauli string patterns in the benchmarks can be 
classified into two categories based on the numbers of non-identity operators in each Pauli string.
As mentioned in Section~\ref{sec:background}, a Pauli string with more non-identity operators on more qubits will in general be converted to a larger circuit block involving more qubits and gates.
The first category includes the molecule Hamiltonians, the random Hamiltonians, and the UCCSD.
In these Hamiltonians, many Pauli strings have non-identity operators on various numbers of qubits (up to all qubits). % in the simulated system).
The second category includes the Ising, Heisenberg, and the selected QAOA benchmarks, of which the Hamiltonians only have Pauli strings with non-identity operators on at most two qubits.
Such a difference in the operator distribution affects the compilation results.

On the FT backend, benchmarks in the first category (molecule and random Hamiltonians) benefit more from the BC optimizations since Pauli strings with more non-identity operators have larger potential in gate cancellation and depth reduction. 
%While 
Benchmarks in the second category (Ising and Heisenberg) cannot benefit from BC since those Pauli strings with only two non-identity operators can only be synthesized in a single way and there is no space BC can explore to further reduce the gate count and circuit depth.
However,
these benchmarks can benefit a lot from DO.
GCO turns out to be inefficient in both gate count and circuit depth for them because GCO cannot create gate count reduction while DO can create additional single-qubit gate reduction opportunities between consecutive layers by putting many small-size blocks in one layer.
On these benchmarks, DO completely outperforms GCO with on average $84.2\%$ circuit depth reduction and $7.5\%$ total gate count reduction.
%For both DO and GCO scheduling, BC can reduce the total gate count and circuit depth on average by around $8\%$ and $22\%$, respectively.
%However, the difference between DO and GCO on these benchmarks are also amortized since more gate cancellation in GCO naturally yields some circuit depth reduction. 
%DO achieves around $7\%$ circuit depth reduction with $9\%$ total gate count increase on average.
Similarly on the SC backend, the BC improvement on the UCCSD benchmarks (first category) is also more significant compared with the QAOA benchmarks (second category) because more gate can be cancelled and more SWAPs in the mapping overhead can be eliminated when the tree sizes are large for Pauli strings with more non-identity operators.

\subsection{Real System Study}
\begin{figure}
    \centering
    \vspace{5pt}
    \includegraphics[width=1.0\columnwidth]{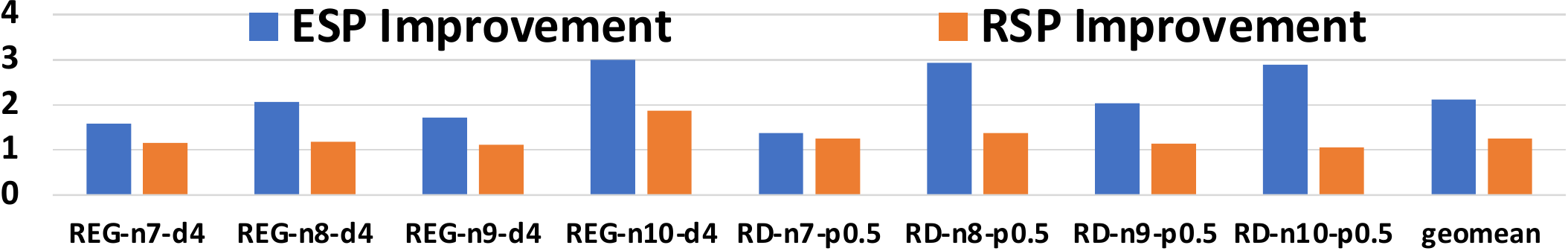}
    \vspace{-15pt}
    \caption{Success Probability Improvement for QAOA on IBM's 16-qubit Melbourne Chip}
    \vspace{-5pt}
    \label{fig:realsystemresult}
\end{figure}

Finally, we evaluate `PH' on IBM's 16-qubit Melbourne chip with 8 QAOA MaxCut programs. 
We generate 4 regular graphs of 7 to 10 nodes with 4 edges per node (`REG-n(7-10)-d4'), and 4 random graphs of 7 to 10 nodes with edge probability 0.5 (`RD-n(7-10)-p0.5').
We prepare 1-level QAOA circuits on these graphs and then optimize the parameters in the simulator.
Those circuits with the optimized parameters are then evaluated on the Melbourne chip (40960 shots per circuit).
The baseline is `Qiskit\_L3' with the Pauli strings ordered by iterating over the adjacency matrix (Qiskit default configuration). 
Figure~\ref{fig:realsystemresult} shows the improvement of the success probability after applying `PH' optimizations. %compared with the baseline.
The `Estimated Success Probability' (ESP), a widely used metric in guiding the compiler optimization~\cite{nishio2020extracting,tannu2019ensemble,murali2019noise}, is a theoretical estimation of the success probability based on the program and the hardware noise model. 
The `Real System Success Probability' (RSP) is the number of trials with correct measurement results divided by the total number of trials when executing on the real machine.
Applying `PH' can improve the ESP by $2.11\times$ on average (up to $3.00\times$) based on the noise model of the tested device, by reducing the CNOT count and circuit depth by 15.1\%  and 36.2\%, respectively on average.
On the real machine, `PH' can improve the RSP by $1.24\times$ on average (up to $1.87\times$).
There is a gap between the results from ESP and RSP because the noise model only provides limited hardware information. 
We expect that the compilation can be further improved with more detailed hardware models.
%and it cannot perfectly characterize the hardware behavior.

\section{Discussion}\label{sec:discussion}

%It would be always desirable to have more effective quantum compiler optimizations to fully exploit the potential of quantum computing. 
%One common approach is to model the hardware more precisely
%(e.g., from coarse-grained gate count~\cite{siraichi2018qubit, li2019tackling,zulehner2019efficient}  to independent non-uniform gate error~\cite{murali2019noise, tannu2019not}, then correlated crosstalk error\cite{murali2020software}, and finally low-level pulse optimizations~\cite{gokhale2020optimized,cheng2020accqoc}). 
%and the compiler can naturally exploit more potential from the hardware with more detailed hardware information.

A recent trend developing in quantum compiler optimizations~\cite{siraichi2018qubit, li2019tackling,zulehner2019efficient,murali2019noise, tannu2019not,murali2020software,gokhale2020optimized,cheng2020accqoc} is to 
%model the hardware more precisely and the compiler can naturally 
exploit more potential from the hardware with more detailed device information.
Different from the innovations in this direction that are mostly driven by the underlying technologies, \myCompilerNameSpace takes another approach which is to enable deeper compiler optimizations by leveraging the algorithmic properties of the high-level quantum programs.
Relatively little attention has been paid to this direction because 
1) it is exceedingly difficult to extract useful high-level semantics from gate sequences, which is the level that most compiler infrastructures today operate at,
and 
2) scalable yet effective static analysis of quantum programs is also very hard as the size of the operation matrices grows exponentially with the number of qubits.
We believe that these are two critical yet difficult open problems in the future development of quantum compiler/software infrastructure since they prevent the compiler from automatically detecting high-level and large-scale optimization opportunities.

\iffalse

\myCompilerNameSpace touches two critical yet difficult open problems in the future development of quantum compiler/software infrastructure.

How to design a with high-level semantics, 

How to derive scalable and effective compiler optimizations at large scale

\fi 

\myCompilerNameSpace tackles these two problems for the \QSK, a widely used subroutine, and thus can benefit the compiler optimization for many quantum algorithms.
In particular, we define a new Pauli IR which can capture the high-level semantics and encode the constraints of all simulation kernels, as far as we know.
The large optimization space of simulation kernels can thus be explored by the compiler automatically, yielding optimization passes that are hard to be implemented in the conventional gate-based representation or ad-hoc algorithm optimizations.
%We then design several new compiler passes, all of which are scalable block-wise circuit transformations since the analysis on Pauli strings can be efficiently handled by classical computers.
%The evaluation in this paper has covered a wide range of \QSK s and 
We expect that \myCompilerNameSpace will continue to benefit future quantum algorithms since quantum simulation has been a long-lived algorithm design principle in the last few decades.

Looking forward, although \myCompilerNameSpace is designed from an algorithmic perspective, it can incorporate those technology-driven optimizations. 
%In this paper, we have supported two different backends with two technology-dependent optimization passes targeting different objectives and hardware constraints.
%In this paper, we have two passes for two backends.
For example, our technology-dependent passes %in this paper
%These passes 
can be further optimized with %once we have %a deeper understanding and 
more comprehensive models of the target %quantum
devices. % and come up with more comprehensive hardware models.
\myCompilerNameSpace can be extended to other technologies
%quantum architectures
(e.g., ion trap~\cite{murali2020architecting,wu2020tilt}, photonics~\cite{Arrazola2021Quantum}) by adding new %optimization
passes.  
It is also possible to make \myCompilerNameSpace more intelligent by automatically managing the passes based on the input program characteristics.
%Currently \myCompilerNameSpace has four passes and 
We have already observed that the different Pauli string patterns can affect the final improvement under different pass configurations as discussed in Section~\ref{sec:evaluation}.
%In the future, more passes can be included to cover more backends, error resources, architectural constraints, and optimization objectives.
How to %automatically 
tune the pass algorithms or derive new suitable passes base on the simulation kernel characteristics
%select the most suitable combination of passes from a pool of compiler passes 
is worth exploring. 
Finally, we believe that the idea of high-level optimization
%quantum 
%algorithmic compiler 
can be extended to other quantum algorithm design techniques (e.g., phase estimation~\cite{nielsen2010quantum}, amplitude amplification~\cite{Brassard1997exact}) and promising quantum application domains (e.g., quantum machine learning~\cite{lloyd2013quantum}). 
%How to design new programming languages to maintain the high-level semantics of these programs and then propose corresponding algorithmic compiler optimizations is still an open problem which can be left as future work.
\iffalse
Finally, the idea of quantum algorithmic compiler can be extended to other promising quantum algorithm domains. 
There are several other important common techniques in quantum algorithm design (e.g., phase estimation~\cite{nielsen2010quantum}, amplitude amplification~\cite{Brassard1997exact}) and promising quantum application domains (e.g., quantum machine learning~\cite{lloyd2013quantum}). 
How to design new programming languages to maintain the high-level semantics of these programs and then propose corresponding algorithmic compiler optimizations is still an open problem which can be left as future work.
\fi
%Finally, as a rapidly developing area, the quantum hardware characteristics and constraints are rapidly evolving. 

%There have been various underlying implementation technologies for near-term QC,  architectures

%photonics~\cite{Arrazola2021Quantum}, ion-trap-based architectures~\cite{murali2020architecting,wu2020tilt}

\section{Related Work}\label{sec:relatedwork}
\myCompilerNameSpace is a compiler framework with a new IR abstraction and deeper optimizations for general quantum simulation kernels.
We first review the program representation and optimizations in quantum compilers. 
Then we discuss existing optimizations for quantum simulation programs. 

\textbf{IR in quantum compilers:} 
%The format of the program representation inside a compiler will affect the optimizations that can be applied. 
Modern classical compilers %usually
employ multiple IRs (e.g., control flow graph, static single assignment) from high level to low level and different optimizations are applied on different IRs. % during compilation.
Today's quantum compilers~\cite{Qiskit,sivarajah2020t,amy2020staq,khammassi2020openql,mccaskey2021mlir}, on the other hand, are mostly built around low-level representations~\cite{cross2017open,smith2016practical,kissinger2019pyzx}, which makes it difficult to extract high-level information about the semantics of the algorithm and discover non-commutative yet semantics-preserving re-orderings. The most recent version of open quantum assembly language (OpenQASM)~\cite{cross2021openqasm} recognizes the need for higher-level semantics such as control, inverse and power operations, but is still incapable of representing Pauli-level semantics which are prevalent in quantum simulation kernels. As we have shown, our Pauli IR can carry high-level semantics through multiple optimization stages, encode all known algorithm constraints, and is compatible with further low-level optimizations by these tools.
\textbf{Quantum compiler optimizations:}
The state-of-the-art quantum compilers~\cite{Qiskit,smith2016practical} usually have multiple passes to execute different optimizations, (e.g., circuit rewriting~\cite{soeken2013white}, gate cancellation~\cite{nam2018automated}, template matching~\cite{maslov2008quantum}, qubit mapping~\cite{murali2019noise}).
These passes applied on the low-level gate sequences usually only rewrite the circuit locally on very few qubits or gates every time and only focus on one optimization objective in each pass.
Different from these optimizations, all passes in \myCompilerNameSpace performance program transformations at a much larger scope in a scalable way and multiple optimization opportunities can be reconciled because the high-level algorithmic information is leveraged.
This makes \myCompilerNameSpace optimizations more effective than simply combining those small-scale single-objective passes. 

\textbf{Optimizations for simulation algorithms:}
%One common technique to optimize the simulation algorithms 
One common optimization technique
is to group the Pauli strings into sets of mutually commutative strings and then apply simultaneous diagonalization~\cite{cowtan2019phase, cowtan2020generic, de2020architecture, van2020circuit}. 
This technique, adopted by $\rm t|ket\rangle$~\cite{sivarajah2020t, cowtan2019phase, cowtan2020generic, de2020architecture}, can simplify the circuit inside each set while the simultaneous diagonalization step introduces substantial overhead before and after the circuit of each set, limiting the overall optimization performance.
%Some other works explore Pauli string scheduling or synthesis but these works are mostly ad hoc, limited to specific algorithms/architectures, and not easy to be generalized to a broader range of programs and employed by a compiler infrastructure. %due to the lack of a good high-level IR.
Some other works~\cite{hastings2015improving, gui2020term, vandaele2021phase, li2021software, shi2019optimized,alam2020circuit,tan2020optimal} explore the simulation program optimization or synthesis but these works are mostly ad-hoc, limited to specific algorithms/architectures, and not easily generalizable to a broader range of programs and employed by a compiler infrastructure. 
\iffalse
\cite{hastings2015improving} and~\cite{gui2020term} reorder the Pauli strings in the electron structure simulation problems for gate cancellation and error suppression.
\cite{vandaele2021phase} optimize the phase-polynomial synthesis on SC architectures where phase-polynomial is a very special class of Pauli strings.
\cite{li2021software} reduces the mapping overhead of UCCSD on a very special SC architecture.
It has also been observed that the CPHASE gates, which can be considered as the $ZZ$ Pauli string simulation, in the QAOA ansatz for unconstrained optimization problems are commutative and they can be freely rescheduled~\cite{shi2019optimized,alam2020circuit,tan2020optimal}. 
%for compiler optimization on SC architectures~\cite{shi2019optimized,alam2020circuit,tan2020optimal}. 
These works are based on the specific two-qubit gate property and are hard to be generalized to other simulation kernels, e.g., even the QAOA for constrained optimization problems~\cite{saleem2020approaches}.
\fi
In \myCompilerName, the Pauli IR's recursive, block-wise structure can support simulation kernels in all related algorithm, as far as we know.
And our optimization algorithms have been shown to be much more effective in the evaluation above.

\section{Conclusion}\label{sec:conclusion}

%In this paper, w
We propose \myCompilerName, an algorithmic quantum compiler targeting the \QSK, a subroutine widely used in many quantum algorithms.
\myCompilerNameSpace enables deep compiler optimizations by defining a new Pauli-string-based IR, which can encode high-level algorithmic information and constraints of many seemingly different quantum algorithms in a unified manner.
All follow-up optimizations in \myCompilerNameSpace operate at a large scope with good scalability and can reconcile multiple optimization opportunities.
\myCompilerNameSpace can be extended to different backends by adding or modifying technology-dependent passes.
Comprehensive experimental results show that \myCompilerNameSpace can significantly outperform state-of-the-art quantum compilers with %more scalable analysis, 
more effective, scalable optimizations and better reconfigurability.

\bibliographystyle{unsrt}
\bibliography{references}

\begin{thebibliography}{10}

\bibitem{feynman1982simulating}
Richard~P Feynman.
\newblock Simulating physics with computers.
\newblock {\em Int. J. Theor. Phys}, 21(6/7), 1982.

\bibitem{lloyd1996universal}
Seth Lloyd.
\newblock Universal quantum simulators.
\newblock {\em Science}, pages 1073--1078, 1996.

\bibitem{abrams1999quantum}
Daniel~S. Abrams and Seth Lloyd.
\newblock Quantum algorithm providing exponential speed increase for finding
  eigenvalues and eigenvectors.
\newblock {\em Phys. Rev. Lett.}, 83:5162--5165, Dec 1999.

\bibitem{harrow2009quantum}
Aram~W. Harrow, Avinatan Hassidim, and Seth Lloyd.
\newblock Quantum algorithm for linear systems of equations.
\newblock {\em Phys. Rev. Lett.}, 103:150502, Oct 2009.

\bibitem{lloyd2014quantum}
Seth Lloyd, Masoud Mohseni, and Patrick Rebentrost.
\newblock Quantum principal component analysis.
\newblock {\em Nature Physics}, 10(9):631--633, 2014.

\bibitem{rebentrost2014quantum}
Patrick Rebentrost, Masoud Mohseni, and Seth Lloyd.
\newblock Quantum support vector machine for big data classification.
\newblock {\em Phys. Rev. Lett.}, 113:130503, Sep 2014.

\bibitem{peruzzo2014variational}
Alberto Peruzzo, Jarrod McClean, Peter Shadbolt, Man-Hong Yung, Xiao-Qi Zhou,
  Peter~J Love, Al{\'a}n Aspuru-Guzik, and Jeremy~L O’brien.
\newblock A variational eigenvalue solver on a photonic quantum processor.
\newblock {\em Nature communications}, 5(1):1--7, 2014.

\bibitem{farhi2014quantum}
Edward Farhi, Jeffrey Goldstone, and Sam Gutmann.
\newblock A quantum approximate optimization algorithm.
\newblock {\em arXiv preprint arXiv:1411.4028}, 2014.

\bibitem{Qiskit}
H{\'e}ctor Abraham, AduOffei, Rochisha Agarwal, Ismail~Yunus Akhalwaya, Gadi
  Aleksandrowicz, Thomas Alexander, Matthew Amy, Eli Arbel, Arijit02, Abraham
  Asfaw, Artur Avkhadiev, Carlos Azaustre, AzizNgoueya, Abhik Banerjee, Aman
  Bansal, Panagiotis Barkoutsos, Ashish Barnawal, George Barron, George~S.
  Barron, Luciano Bello, Yael Ben-Haim, Daniel Bevenius, Arjun Bhobe, Lev~S.
  Bishop, Carsten Blank, Sorin Bolos, Samuel Bosch, Brandon, Sergey Bravyi,
  Bryce-Fuller, David Bucher, Artemiy Burov, Fran Cabrera, Padraic Calpin,
  Lauren Capelluto, Jorge Carballo, Gin{\'e}s Carrascal, Adrian Chen, Chun-Fu
  Chen, Edward Chen, Jielun~(Chris) Chen, Richard Chen, Jerry~M. Chow, Spencer
  Churchill, Christian Claus, Christian Clauss, Romilly Cocking, Filipe Correa,
  Abigail~J. Cross, Andrew~W. Cross, Simon Cross, Juan Cruz-Benito, Chris
  Culver, Antonio~D. C{\'o}rcoles-Gonzales, Sean Dague, Tareq~El Dandachi,
  Marcus Daniels, Matthieu Dartiailh, DavideFrr, Abd{\'o}n~Rodr{\'\i}guez
  Davila, Anton Dekusar, Delton Ding, Jun Doi, Eric Drechsler, Drew, Eugene
  Dumitrescu, Karel Dumon, Ivan Duran, Kareem EL-Safty, Eric Eastman, Grant
  Eberle, Pieter Eendebak, Daniel Egger, Mark Everitt, Paco~Mart{\'\i}n
  Fern{\'a}ndez, Axel~Hern{\'a}ndez Ferrera, Romain Fouilland,
  FranckChevallier, Albert Frisch, Andreas Fuhrer, Bryce Fuller, MELVIN GEORGE,
  Julien Gacon, Borja~Godoy Gago, Claudio Gambella, Jay~M. Gambetta, Adhisha
  Gammanpila, Luis Garcia, Tanya Garg, Shelly Garion, Austin Gilliam, Aditya
  Giridharan, Juan Gomez-Mosquera, Gonzalo, Salvador de~la Puente~Gonz{\'a}lez,
  Jesse Gorzinski, Ian Gould, Donny Greenberg, Dmitry Grinko, Wen Guan, John~A.
  Gunnels, Mikael Haglund, Isabel Haide, Ikko Hamamura, Omar~Costa Hamido,
  Frank Harkins, Vojtech Havlicek, Joe Hellmers, {\L}ukasz Herok, Stefan
  Hillmich, Hiroshi Horii, Connor Howington, Shaohan Hu, Wei Hu, Junye Huang,
  Rolf Huisman, Haruki Imai, Takashi Imamichi, Kazuaki Ishizaki, Raban Iten,
  Toshinari Itoko, JamesSeaward, Ali Javadi, Ali Javadi-Abhari, Wahaj Javed,
  Jessica, Madhav Jivrajani, Kiran Johns, Scott Johnstun, Jonathan-Shoemaker,
  Vismai K, Tal Kachmann, Akshay Kale, Naoki Kanazawa, Kang-Bae, Anton
  Karazeev, Paul Kassebaum, Josh Kelso, Spencer King, Knabberjoe, Yuri
  Kobayashi, Arseny Kovyrshin, Rajiv Krishnakumar, Vivek Krishnan, Kevin
  Krsulich, Prasad Kumkar, Gawel Kus, Ryan LaRose, Enrique Lacal, Rapha{\"e}l
  Lambert, John Lapeyre, Joe Latone, Scott Lawrence, Christina Lee, Gushu Li,
  Dennis Liu, Peng Liu, Yunho Maeng, Kahan Majmudar, Aleksei Malyshev, Joshua
  Manela, Jakub Marecek, Manoel Marques, Dmitri Maslov, Dolph Mathews, Atsushi
  Matsuo, Douglas~T. McClure, Cameron McGarry, David McKay, Dan McPherson,
  Srujan Meesala, Thomas Metcalfe, Martin Mevissen, Andrew Meyer, Antonio
  Mezzacapo, Rohit Midha, Zlatko Minev, Abby Mitchell, Nikolaj Moll, Jhon
  Montanez, Gabriel Monteiro, Michael~Duane Mooring, Renier Morales, Niall
  Moran, Mario Motta, MrF, Prakash Murali, Jan M{\"u}ggenburg, David Nadlinger,
  Ken Nakanishi, Giacomo Nannicini, Paul Nation, Edwin Navarro, Yehuda Naveh,
  Scott~Wyman Neagle, Patrick Neuweiler, Johan Nicander, Pradeep Niroula, Hassi
  Norlen, NuoWenLei, Lee~James O'Riordan, Oluwatobi Ogunbayo, Pauline
  Ollitrault, Raul Otaolea, Steven Oud, Dan Padilha, Hanhee Paik, Soham Pal,
  Yuchen Pang, Vincent~R. Pascuzzi, Simone Perriello, Anna Phan, Francesco
  Piro, Marco Pistoia, Christophe Piveteau, Pierre Pocreau, Alejandro
  Pozas-Kerstjens, Milos Prokop, Viktor Prutyanov, Daniel Puzzuoli, Jes{\'u}s
  P{\'e}rez, Quintiii, Rafey~Iqbal Rahman, Arun Raja, Nipun Ramagiri, Anirudh
  Rao, Rudy Raymond, Rafael Mart{\'\i}n-Cuevas Redondo, Max Reuter, Julia Rice,
  Matt Riedemann, Marcello~La Rocca, Diego~M. Rodr{\'\i}guez, RohithKarur, Max
  Rossmannek, Mingi Ryu, Tharrmashastha SAPV, SamFerracin, Martin Sandberg,
  Hirmay Sandesara, Ritvik Sapra, Hayk Sargsyan, Aniruddha Sarkar, Ninad
  Sathaye, Bruno Schmitt, Chris Schnabel, Zachary Schoenfeld, Travis~L.
  Scholten, Eddie Schoute, Joachim Schwarm, Ismael~Faro Sertage, Kanav Setia,
  Nathan Shammah, Yunong Shi, Adenilton Silva, Andrea Simonetto, Nick
  Singstock, Yukio Siraichi, Iskandar Sitdikov, Seyon Sivarajah, Magnus~Berg
  Sletfjerding, John~A. Smolin, Mathias Soeken, Igor~Olegovich Sokolov, Igor
  Sokolov, SooluThomas, Starfish, Dominik Steenken, Matt Stypulkoski, Shaojun
  Sun, Kevin~J. Sung, Hitomi Takahashi, Tanvesh Takawale, Ivano Tavernelli,
  Charles Taylor, Pete Taylour, Soolu Thomas, Mathieu Tillet, Maddy Tod,
  Miroslav Tomasik, Enrique de~la Torre, Kenso Trabing, Matthew Treinish,
  TrishaPe, Davindra Tulsi, Wes Turner, Yotam Vaknin, Carmen~Recio Valcarce,
  Francois Varchon, Almudena~Carrera Vazquez, Victor Villar, Desiree Vogt-Lee,
  Christophe Vuillot, James Weaver, Johannes Weidenfeller, Rafal Wieczorek,
  Jonathan~A. Wildstrom, Erick Winston, Jack~J. Woehr, Stefan Woerner, Ryan
  Woo, Christopher~J. Wood, Ryan Wood, Stephen Wood, Steve Wood, James Wootton,
  Daniyar Yeralin, David Yonge-Mallo, Richard Young, Jessie Yu, Christopher
  Zachow, Laura Zdanski, Helena Zhang, Christa Zoufal, and Mantas
  {\v{C}}epulkovskis.
\newblock Qiskit: An open-source framework for quantum computing, 2019.

\bibitem{smith2020open}
Robert~S Smith, Eric~C Peterson, Mark~G Skilbeck, and Erik~J Davis.
\newblock An open-source, industrial-strength optimizing compiler for quantum
  programs.
\newblock {\em Quantum Science and Technology}, 5(4):044001, 2020.

\bibitem{sivarajah2020t}
Seyon Sivarajah, Silas Dilkes, Alexander Cowtan, Will Simmons, Alec Edgington,
  and Ross Duncan.
\newblock $\rm t|ket\rangle$: a retargetable compiler for nisq devices.
\newblock {\em Quantum Science and Technology}, 6(1):014003, 2020.

\bibitem{hastings2015improving}
Matthew~B Hastings, Dave Wecker, Bela Bauer, and Matthias Troyer.
\newblock Improving quantum algorithms for quantum chemistry.
\newblock {\em Quantum Information \& Computation}, 15(1-2):1--21, 2015.

\bibitem{gui2020term}
Kaiwen Gui, Teague Tomesh, Pranav Gokhale, Yunong Shi, Frederic~T Chong,
  Margaret Martonosi, and Martin Suchara.
\newblock Term grouping and travelling salesperson for digital quantum
  simulation.
\newblock {\em arXiv preprint arXiv:2001.05983}, 2020.

\bibitem{van2020circuit}
Ewout van~den Berg and Kristan Temme.
\newblock Circuit optimization of hamiltonian simulation by simultaneous
  diagonalization of pauli clusters.
\newblock {\em Quantum}, 4:322, 2020.

\bibitem{cowtan2019phase}
Alexander Cowtan, Silas Dilkes, Ross Duncan, Will Simmons, and Seyon Sivarajah.
\newblock Phase gadget synthesis for shallow circuits.
\newblock {\em arXiv preprint arXiv:1906.01734}, 2019.

\bibitem{cowtan2020generic}
Alexander Cowtan, Will Simmons, and Ross Duncan.
\newblock A generic compilation strategy for the unitary coupled cluster
  ansatz.
\newblock {\em arXiv preprint arXiv:2007.10515}, 2020.

\bibitem{de2020architecture}
Arianne Meijer-van de~Griend and Ross Duncan.
\newblock Architecture-aware synthesis of phase polynomials for nisq devices.
\newblock {\em arXiv preprint arXiv:2004.06052}, 2020.

\bibitem{vandaele2021phase}
Vivien Vandaele, Simon Martiel, and Timoth{\'e}e~Goubault de~Brugi{\`e}re.
\newblock Phase polynomials synthesis algorithms for nisq architectures and
  beyond.
\newblock {\em arXiv preprint arXiv:2104.00934}, 2021.

\bibitem{shi2019optimized}
Yunong Shi, Nelson Leung, Pranav Gokhale, Zane Rossi, David~I Schuster, Henry
  Hoffmann, and Frederic~T Chong.
\newblock Optimized compilation of aggregated instructions for realistic
  quantum computers.
\newblock In {\em Proceedings of the Twenty-Fourth International Conference on
  Architectural Support for Programming Languages and Operating Systems}, pages
  1031--1044, 2019.

\bibitem{alam2020circuit}
Mahabubul Alam, Abdullah Ash-Saki, and Swaroop Ghosh.
\newblock Circuit compilation methodologies for quantum approximate
  optimization algorithm.
\newblock In {\em 53rd Annual IEEE/ACM International Symposium on
  Microarchitecture (MICRO)}, pages 215--228. IEEE, 2020.

\bibitem{tan2020optimal}
Bochen Tan and Jason Cong.
\newblock Optimal layout synthesis for quantum computing.
\newblock In {\em 2020 IEEE/ACM International Conference On Computer Aided
  Design (ICCAD)}, pages 1--9. IEEE, 2020.

\bibitem{li2021software}
Gushu Li, Yunong Shi, and Ali Javadi-Abhari.
\newblock Software-hardware co-optimization for computational chemistry on
  superconducting quantum processors.
\newblock {\em arXiv preprint arXiv:2105.07127}, 2021.

\bibitem{lao20212qan}
Lingling Lao and Dan Browne.
\newblock 2qan: A quantum compiler for 2-local qubit hamiltonian simulation
  algorithms, 2021.

\bibitem{soeken2013white}
Mathias Soeken and Michael~Kirkedal Thomsen.
\newblock White dots do matter: rewriting reversible logic circuits.
\newblock In {\em International Conference on Reversible Computation}, pages
  196--208. Springer, 2013.

\bibitem{nam2018automated}
Yunseong Nam, Neil~J Ross, Yuan Su, Andrew~M Childs, and Dmitri Maslov.
\newblock Automated optimization of large quantum circuits with continuous
  parameters.
\newblock {\em npj Quantum Information}, 4(1):1--12, 2018.

\bibitem{maslov2008quantum}
Dmitri Maslov, Gerhard~W Dueck, D~Michael Miller, and Camille Negrevergne.
\newblock Quantum circuit simplification and level compaction.
\newblock {\em IEEE Transactions on Computer-Aided Design of Integrated
  Circuits and Systems}, 27(3):436--444, 2008.

\bibitem{murali2019noise}
Prakash Murali, Jonathan~M Baker, Ali Javadi-Abhari, Frederic~T Chong, and
  Margaret Martonosi.
\newblock Noise-adaptive compiler mappings for noisy intermediate-scale quantum
  computers.
\newblock In {\em Proceedings of the Twenty-Fourth International Conference on
  Architectural Support for Programming Languages and Operating Systems}, pages
  1015--1029, 2019.

\bibitem{alam2020efficient}
Mahabubul Alam, Abdullah Ash-Saki, and Swaroop Ghosh.
\newblock An efficient circuit compilation flow for quantum approximate
  optimization algorithm.
\newblock In {\em 2020 57th ACM/IEEE Design Automation Conference (DAC)}, pages
  1--6. IEEE, 2020.

\bibitem{alam2020noise}
Mahabubul Alam, Abdullah Ash-Saki, Junde Li, Anupam Chattopadhyay, and Swaroop
  Ghosh.
\newblock Noise resilient compilation policies for quantum approximate
  optimization algorithm.
\newblock In {\em Proceedings of the 39th International Conference on
  Computer-Aided Design}, pages 1--7, 2020.

\bibitem{nielsen2010quantum}
Michael~A Nielsen and Isaac~L Chuang.
\newblock Quantum computation and quantum information.
\newblock {\em Quantum Computation and Quantum Information, by Michael A.
  Nielsen, Isaac L. Chuang, Cambridge, UK: Cambridge University Press, 2010},
  2010.

\bibitem{trotter1959product}
Hale~F Trotter.
\newblock On the product of semi-groups of operators.
\newblock {\em Proceedings of the American Mathematical Society},
  10(4):545--551, 1959.

\bibitem{cerezo2020variational}
Marco Cerezo, Andrew Arrasmith, Ryan Babbush, Simon~C Benjamin, Suguru Endo,
  Keisuke Fujii, Jarrod~R McClean, Kosuke Mitarai, Xiao Yuan, Lukasz Cincio,
  and Patrick~J. Coles.
\newblock Variational quantum algorithms.
\newblock {\em arXiv preprint arXiv:2012.09265}, 2020.

\bibitem{saleem2020approaches}
Zain~H Saleem, Bilal Tariq, and Martin Suchara.
\newblock Approaches to constrained quantum approximate optimization.
\newblock {\em arXiv preprint arXiv:2010.06660}, 2020.

\bibitem{gard2020efficient}
Bryan~T. Gard, Linghua Zhu, George~S. Barron, Nicholas~J. Mayhall, Sophia~E.
  Economou, and Edwin Barnes.
\newblock Efficient symmetry-preserving state preparation circuits for the
  variational quantum eigensolver algorithm.
\newblock {\em npj Quantum Information}, 6(1):10, Jan 2020.

\bibitem{Tranter2018comparison}
Andrew Tranter, Peter~J. Love, Florian Mintert, and Peter~V. Coveney.
\newblock A comparison of the bravyi–kitaev and jordan–wigner
  transformations for the quantum simulation of quantum chemistry.
\newblock {\em Journal of Chemical Theory and Computation}, 14(11):5617--5630,
  2018.
\newblock PMID: 30189144.

\bibitem{fowler2012surface}
Austin~G. Fowler, Matteo Mariantoni, John~M. Martinis, and Andrew~N. Cleland.
\newblock Surface codes: Towards practical large-scale quantum computation.
\newblock {\em Phys. Rev. A}, 86:032324, Sep 2012.

\bibitem{chamberland2020topological}
Christopher Chamberland, Guanyu Zhu, Theodore~J. Yoder, Jared~B. Hertzberg, and
  Andrew~W. Cross.
\newblock Topological and subsystem codes on low-degree graphs with flag
  qubits.
\newblock {\em Phys. Rev. X}, 10:011022, Jan 2020.

\bibitem{maslov2016optimal}
Dmitri Maslov.
\newblock Optimal and asymptotically optimal nct reversible circuits by the
  gate types.
\newblock {\em arXiv preprint arXiv:1602.02627}, 2016.

\bibitem{PYSCF}
Qiming Sun, Timothy~C. Berkelbach, Nick~S. Blunt, George~H. Booth, Sheng Guo,
  Zhendong Li, Junzi Liu, James~D. McClain, Elvira~R. Sayfutyarova, Sandeep
  Sharma, Sebastian Wouters, and Garnet~Kin‐Lic Chan.
\newblock Pyscf: the python‐based simulations of chemistry framework, 2017.

\bibitem{nishio2020extracting}
Shin Nishio, Yulu Pan, Takahiko Satoh, Hideharu Amano, and Rodney~Van Meter.
\newblock Extracting success from ibm’s 20-qubit machines using error-aware
  compilation.
\newblock {\em J. Emerg. Technol. Comput. Syst.}, 16(3), May 2020.

\bibitem{tannu2019ensemble}
Swamit~S Tannu and Moinuddin Qureshi.
\newblock Ensemble of diverse mappings: Improving reliability of quantum
  computers by orchestrating dissimilar mistakes.
\newblock In {\em Proceedings of the 52nd Annual IEEE/ACM International
  Symposium on Microarchitecture}, pages 253--265, 2019.

\bibitem{siraichi2018qubit}
Marcos~Yukio Siraichi, Vin{\'\i}cius Fernandes~dos Santos, Sylvain Collange,
  and Fernando Magno~Quint{\~a}o Pereira.
\newblock Qubit allocation.
\newblock In {\em Proceedings of the 2018 International Symposium on Code
  Generation and Optimization}, pages 113--125, 2018.

\bibitem{li2019tackling}
Gushu Li, Yufei Ding, and Yuan Xie.
\newblock Tackling the qubit mapping problem for nisq-era quantum devices.
\newblock In {\em Proceedings of the Twenty-Fourth International Conference on
  Architectural Support for Programming Languages and Operating Systems}, pages
  1001--1014, 2019.

\bibitem{zulehner2019efficient}
A.~{Zulehner}, A.~{Paler}, and R.~{Wille}.
\newblock An efficient methodology for mapping quantum circuits to the ibm qx
  architectures.
\newblock {\em IEEE Transactions on Computer-Aided Design of Integrated
  Circuits and Systems}, 38(7):1226--1236, 2019.

\bibitem{tannu2019not}
Swamit~S Tannu and Moinuddin~K Qureshi.
\newblock Not all qubits are created equal: a case for variability-aware
  policies for nisq-era quantum computers.
\newblock In {\em Proceedings of the Twenty-Fourth International Conference on
  Architectural Support for Programming Languages and Operating Systems}, pages
  987--999, 2019.

\bibitem{murali2020software}
Prakash Murali, David~C McKay, Margaret Martonosi, and Ali Javadi-Abhari.
\newblock Software mitigation of crosstalk on noisy intermediate-scale quantum
  computers.
\newblock In {\em Proceedings of the Twenty-Fifth International Conference on
  Architectural Support for Programming Languages and Operating Systems}, pages
  1001--1016, 2020.

\bibitem{gokhale2020optimized}
P.~{Gokhale}, A.~{Javadi-Abhari}, N.~{Earnest}, Y.~{Shi}, and F.~T. {Chong}.
\newblock Optimized quantum compilation for near-term algorithms with
  openpulse.
\newblock In {\em 53rd Annual IEEE/ACM International Symposium on
  Microarchitecture (MICRO)}, pages 186--200, 2020.

\bibitem{cheng2020accqoc}
J.~{Cheng}, H.~{Deng}, and X.~{Qia}.
\newblock Accqoc: Accelerating quantum optimal control based pulse generation.
\newblock In {\em ACM/IEEE 47th Annual International Symposium on Computer
  Architecture (ISCA)}, pages 543--555, 2020.

\bibitem{murali2020architecting}
Prakash Murali, Dripto~M Debroy, Kenneth~R Brown, and Margaret Martonosi.
\newblock Architecting noisy intermediate-scale trapped ion quantum computers.
\newblock In {\em ACM/IEEE 47th Annual International Symposium on Computer
  Architecture (ISCA)}, pages 529--542. IEEE, 2020.

\bibitem{wu2020tilt}
Xin-Chuan Wu, Dripto~M Debroy, Yongshan Ding, Jonathan~M Baker, Yuri Alexeev,
  Kenneth~R Brown, and Frederic~T Chong.
\newblock Tilt: Achieving higher fidelity on a trapped-ion linear-tape quantum
  computing architecture.
\newblock {\em arXiv preprint arXiv:2010.15876}, 2020.

\bibitem{Arrazola2021Quantum}
J.~M. Arrazola, V.~Bergholm, K.~Br{\'a}dler, T.~R. Bromley, M.~J. Collins,
  I.~Dhand, A.~Fumagalli, T.~Gerrits, A.~Goussev, L.~G. Helt, J.~Hundal,
  T.~Isacsson, R.~B. Israel, J.~Izaac, S.~Jahangiri, R.~Janik, N.~Killoran,
  S.~P. Kumar, J.~Lavoie, A.~E. Lita, D.~H. Mahler, M.~Menotti, B.~Morrison,
  S.~W. Nam, L.~Neuhaus, H.~Y. Qi, N.~Quesada, A.~Repingon, K.~K. Sabapathy,
  M.~Schuld, D.~Su, J.~Swinarton, A.~Sz{\'a}va, K.~Tan, P.~Tan, V.~D. Vaidya,
  Z.~Vernon, Z.~Zabaneh, and Y.~Zhang.
\newblock Quantum circuits with many photons on a programmable nanophotonic
  chip.
\newblock {\em Nature}, 591(7848):54--60, Mar 2021.

\bibitem{Brassard1997exact}
G.~{Brassard} and P.~{Hoyer}.
\newblock An exact quantum polynomial-time algorithm for simon's problem.
\newblock In {\em Proceedings of the Fifth Israeli Symposium on Theory of
  Computing and Systems}, pages 12--23, 1997.

\bibitem{lloyd2013quantum}
Seth Lloyd, Masoud Mohseni, and Patrick Rebentrost.
\newblock Quantum algorithms for supervised and unsupervised machine learning.
\newblock {\em arXiv preprint arXiv:1307.0411}, 2013.

\bibitem{amy2020staq}
Matthew Amy and Vlad Gheorghiu.
\newblock staq—a full-stack quantum processing toolkit.
\newblock {\em Quantum Science and Technology}, 5(3):034016, 2020.

\bibitem{khammassi2020openql}
Nader Khammassi, Imran Ashraf, J~v Someren, Razvan Nane, AM~Krol, M~Adriaan
  Rol, L~Lao, Koen Bertels, and Carmen~G Almudever.
\newblock Openql: A portable quantum programming framework for quantum
  accelerators.
\newblock {\em arXiv preprint arXiv:2005.13283}, 2020.

\bibitem{mccaskey2021mlir}
Alexander McCaskey and Thien Nguyen.
\newblock A mlir dialect for quantum assembly languages.
\newblock {\em arXiv preprint arXiv:2101.11365}, 2021.

\bibitem{cross2017open}
Andrew~W Cross, Lev~S Bishop, John~A Smolin, and Jay~M Gambetta.
\newblock Open quantum assembly language.
\newblock {\em arXiv preprint arXiv:1707.03429}, 2017.

\bibitem{smith2016practical}
Robert~S Smith, Michael~J Curtis, and William~J Zeng.
\newblock A practical quantum instruction set architecture.
\newblock {\em arXiv preprint arXiv:1608.03355}, 2016.

\bibitem{kissinger2019pyzx}
Aleks Kissinger and John van~de Wetering.
\newblock Pyzx: Large scale automated diagrammatic reasoning.
\newblock {\em arXiv preprint arXiv:1904.04735}, 2019.

\bibitem{cross2021openqasm}
Andrew~W Cross, Ali Javadi-Abhari, Thomas Alexander, Niel de~Beaudrap, Lev~S
  Bishop, Steven Heidel, Colm~A Ryan, John Smolin, Jay~M Gambetta, and Blake~R
  Johnson.
\newblock Openqasm 3: A broader and deeper quantum assembly language.
\newblock {\em arXiv preprint arXiv:2104.14722}, 2021.

\end{thebibliography}

\end{document}